\documentclass[aps,pra,twocolumn,superscriptaddress,showpacs,amsmath,amssymb,nofootinbib]{revtex4}
\usepackage{multirow}

\usepackage{graphicx}   % Include figure files
\usepackage{xspace}     % Needed for \xspace command
\usepackage{array}
\usepackage{color}

\newcommand{\xmath}[1]{\ensuremath{#1}\xspace}
\newcommand{\mrm}  [1]{\mathrm{#1}}
\newcommand{\mbf}  [1]{\mathbf{#1}}
\newcommand{\SSzero}{\xmath{\mrm{ {}^1S_0 }}}       %1S0 : LS-coupling
\newcommand{\TDone} {\xmath{\mrm{ {}^3D_1 }}}       %3D1
\newcommand{\TPtwo} {\xmath{\mrm{ {}^3P_2 }}}       %3P2
\newcommand{\TPone} {\xmath{\mrm{ {}^3P_1 }}}       %3P1
\newcommand{\TPzero}{\xmath{\mrm{ {}^3P_0 }}}       %3P0
\newcommand{\SPone} {\xmath{\mrm{ {}^1P_1 }}}       %1P1
\newcommand{\SSZero}{\xmath{\mrm{ 6s^2 }\:\SSzero }}%1S0 : JJ-coupling
\newcommand{\TDOne} {\xmath{\mrm{ 5d6s }\:\TDone  }}%3D1
\newcommand{\Energy}{\mathcal{E}}
\newcommand{\EA}  {\Energy_{\mrm{A}}}
\newcommand{\EZ}  {\Energy_{\mrm{Z}}}

\newcommand{\Edc}{\Energy_{\mrm{dc}}}
\newcommand{\Eac}{\Energy_{\mrm{ac}}}
\newcommand{\Hatomic}{H_{\mathrm{A}}}
\newcommand{\HZeeman}{H_{\mathrm{Z}}}

\newcommand{\Hac} {H_{\mathrm{ac}}}
\newcommand{\Hdc} {H_{\mathrm{dc}}}
\newcommand{\elight}{\xmath{E}}
\newcommand{\klight}{\xmath{k}}
\newcommand{\bfield}{\xmath{B}}
\newcommand{\efield}{\xmath{E}_{\mrm{dc}}}
\newcommand{\Elight}{\xmath{\mbf{E}}}
\newcommand{\Klight}{\xmath{\mbf{\klight}}}
\newcommand{\Bfield}{\xmath{\mbf{\bfield}}}
\newcommand{\Efield}{\xmath{\mbf{E}_{\mrm{dc}}}}
\newcommand{\edipole}{\xmath{d}}
\newcommand{\mdipole}{\xmath{\mu}}
\newcommand{\Edipole}{\xmath{\mbf{\edipole}}}
\newcommand{\Mdipole}{\xmath{\boldsymbol{\mdipole}}}
\newcommand{\xhat}{\xmath{\hat{\mathbf{x}}}}
\newcommand{\yhat}{\xmath{\hat{\mathbf{y}}}}
\newcommand{\zhat}{\xmath{\hat{\mathbf{z}}}}

\newcommand{\ehat}{\xmath{\boldsymbol{\epsilon}}}
\newcommand{\rvec}{\xmath{\mbf{r}}}
\newcommand{\bra}[1]{\langle#1|}
\newcommand{\ket}[1]{|#1\rangle}

\newcommand{\kq}[3]{\xmath{#1^{(#2)}_{#3}}}
\newcommand{\reducedME}[4]{\xmath{(#1||#2^{(#3)}||#4)}}
\newcommand{\adc}{\alpha^{\mrm{dc}}}
\newcommand{\aac}{\alpha}
\newcommand{\vT}{\xmath{v_\mathrm{T}}}
\newcommand{\vC}{\xmath{v_\mathrm{C}}}
\newcommand{\kB}  {\xmath{k_B}}
\newcommand{\muB} {\xmath{\mu_0}}

\newcommand{\rbeam}{r_{\mrm{b}}}
\newcommand{\dind}{d_{\mrm{in}}}
\newcommand{\woven}{w_{\mathrm{T}}}
\newcommand{\wcoll}{w_{\mathrm{L}}}

\newcommand{\Gammap}{\Omega}

\newcommand{\ti}{\tau_{\mrm{i}}}
\newcommand{\tf}{\tau_{\mrm{f}}}
\newcommand{\omegaD}{\omega_{\mrm{D}}}
\renewcommand{\Re}{\mathrm{Re}}
\renewcommand{\Im}{\mathrm{Im}}

\newcommand{\Am}{A_{\mrm{m}}}
\newcommand{\omegam}{\omega_{\mrm{m}}}

\newcommand{\drdf}[1]{#1}

\begin{document}
\title{Measurement of dynamic Stark polarizabilities by analyzing spectral lineshapes of forbidden
transitions}
\author{D.~R.~Dounas-Frazer}
\email{drdf@berkeley.edu}
\author{K.~Tsigutkin}
\author{A.~Family}
\affiliation{Department of Physics, University of California at
Berkeley, Berkeley, CA 94720-7300}
\author{D.~Budker}
\affiliation{Department of Physics, University of California at
Berkeley, Berkeley, CA 94720-7300} \affiliation{Nuclear Science
Division, Lawrence Berkeley National Laboratory, Berkeley,
California 94720}
\date{\today}

\begin{abstract}
We present a measurement of the dynamic scalar and tensor
polarizabilities of the excited state $\ket{\TDOne}$ in atomic
ytterbium. The polarizabilities were measured by analyzing the
spectral lineshape of the 408-nm $\SSZero\rightarrow\TDOne$
transition driven by a standing wave of resonant light in the
presence of static electric and magnetic fields. Due to the
interaction of atoms with the standing wave, the lineshape has a
characteristic polarizability-dependent distortion. A theoretical
model was used to simulate the lineshape and determine a combination
of the polarizabilities of the ground and excited states by fitting
the model to experimental data. This combination was measured with a
13\% uncertainty, only 3\% of which is due to uncertainty in the
simulation and fitting procedure. The scalar and tensor
polarizabilities of the state $\ket{\TDOne}$ were measured for the
first time by comparing two different combinations of
polarizabilities. We show that this technique can be applied to
similar atomic systems.

\end{abstract}

\pacs{32.90.+a, 32.70.Jz, 32.60.+i}

\maketitle

% =====================================
% INTRODUCTION
% =====================================
\section{Introduction}\label{sec:Intro}
Static (dc) and dynamic (ac) electric dipole polarizabilities
determine the response of neutral particles to applied electric
fields.  { They are related to a host of atomic and molecular
quantities, including the dielectric constant, refractive index, and
Stark
shift~\cite{ref:Autler1955,ref:Angel1968,ref:Bonin1994,ref:Bonin1997},
and are an important consideration for many current atomic,
molecular, and optical physics experiments~\cite{ref:Mitroy2010}. }
For example, polarizabilities play a vital role in the production of
light traps for quantum information processing
applications~\cite{ref:Ye2008}.  {In the context of optical atomic
clocks, Stark shifts constitute an important systematic effect that
must be controlled~\cite{ref:Udem2005, ref:Takamoto2005,
ref:Barber2008, ref:Chou2010}.}  Similarly, Stark shifts also
contribute to systematic effects in atomic parity violation (APV)
measurements~\cite{ref:Wieman1987, ref:Wood1999, ref:Tsigutkin2009,
ref:Tsigutkin2010}. Hence the determination of polarizabilities is a
priority for high-precision atomic physics.

Present experiments typically rely on theoretical calculations of
electric dipole
polarizabilities~\cite{ref:Dzuba2010,ref:Safronova2010,ref:Beck2010}.
Several methods for measuring polarizabilities also exist. Early
schemes involved the deflection of atoms in an inhomogeneous
electric field~\cite{ref:Bonin1994}.   {More recent techniques
include absolute frequency measurements~\cite{ref:Barber2008}, atom
interferometry~\cite{ref:Morinaga1993, ref:Ekstrom1995,
ref:Deissler2008}, and a technique that uses light
force~\cite{ref:Kadar-Kallen1992, ref:Kadar-Kallen1994}.}  However,
these methods  {typically} provide information about the
polarizability of an atom in its \emph{ground}
state~\cite{ref:Mitroy2010}. Therefore, they are inappropriate for
high-precision experiments where the polarizabilities of
\emph{excited} states are relevant.

As part of an ongoing investigation of parity violation in atomic
ytterbium (Yb)~\cite{ref:Tsigutkin2009,ref:Tsigutkin2010}, a scheme
for measuring a combination of polarizabilities of the ground and an
excited state of Yb was developed~\cite{ref:Stalnaker2006}. The
scheme involves the simulation and measurement of the spectral
lineshape of a forbidden electric dipole transition driven by a
standing wave of light in the presence of a dc electric field.  Due
to the standing wave, the ac Stark shifts of the upper and lower
states introduce a polarizability-dependent distortion in the
lineshape, a phenomenon which was first observed and characterized
during a search for APV in cesium~\cite{ref:Wieman1987}. The
difference of polarizabilities of the two states is treated as a
variable parameter in the simulation and is measured by fitting the
simulated lineshape to experimental data. We call this scheme the
Lineshape Simulation Method (LSM).  {The LSM can be generalized to
an arbitrary atomic species.}

In this paper, we present the next generation of the LSM.  The
numerical procedures accommodate for a broad domain of values of
input parameters, \emph{e.g.}, the intensity of the standing wave.
In addition, the independent dimensionless parameters that determine
the lineshape have been explicitly identified, thus facilitating
error analysis. \drdf{In general, the LSM is compatible with a
variety of atomic species and field geometries.} The LSM is
sensitive to the \emph{difference} of polarizabilities of the atom
in its ground and excited states. Nevertheless, this method
\drdf{may yield} unambiguous measurements of the vector and tensor
components of the excited state polarizability, as we will show. In
this sense, we present a versatile method for measuring the
excited-state polarizabilities of atoms.

We demonstrate the LSM using the 408-nm
\mbox{$\SSZero\rightarrow\TDOne$} transition in atomic Yb. Whereas
the previous results~\cite{ref:Stalnaker2006} were obtained in the
absence of a magnetic field, the present work uses a magnetic field
to isolate Zeeman sublevels of the excited state. The ac Stark
shifts of the sublevels are characterized by different combinations
of scalar, vector, and tensor polarizabilities. This approach allows
for the unambiguous determination of these polarizabilities. The ac
scalar and tensor polarizabilities of the excited state
$\ket{\TDOne}$ in Yb are measured for the first time. Due to
improvements in the experimental apparatus, the signal-to-noise
ratio of the observed lineshape is an order of magnitude larger than
for the previous implementation. In the present work, the
statistical error introduced by the LSM is negligible compared to
the systematic uncertainty of the experiment.

This paper is organized as follows. In Section
\ref{sec:AtomicSystem}, we introduce our conventions for the
polarizabilities and the Stark shift. The theoretical model,
numerical procedure, and results of the simulation are discussed in
Section~\ref{sec:Lineshape}. In Section~\ref{sec:App} we apply the
LSM to the Yb system and present the results.  Finally, a summary of
the results and an outlook for future experiments \drdf{are} given
in Section~\ref{sec:Outlook}.

% =====================================
% ATOMIC SYSTEM
% =====================================
\section{Atomic system}\label{sec:AtomicSystem}
\drdf{Throughout this work, we consider a Stark-induced transition
between two atomic states of the same parity. The transition is
induced by applying a uniform dc electric field \Efield. We assume
that the transition is driven by a standing wave of light formed by
two counter-propagating waves with the same polarization traveling
in the $\pm\Klight$ directions. In this case, the electric field of
the light is given by}
\begin{equation}\label{eq:StandingWave}
\Elight(\rvec,t) = \elight(\rvec)\cos(\omega t)\,\ehat,
\end{equation}
where
\begin{equation}\label{eq:E}
 E(\rvec) =
E_0\cos(\Klight\cdot\rvec)e^{-r_\perp^2/\rbeam^2}.
\end{equation}
\drdf{Here $E_0$, $\Klight$, $\omega$, and $\ehat$ are the
amplitude, wave-vector, angular frequency, and polarization of the
electric field, respectively, $\rbeam$ is the radius of the standing
wave, and $r_\perp=|\rvec-\hat{\Klight}(\hat{\Klight}\cdot\rvec)|$
is the perpendicular distance from the center of the standing wave.
In addition to these parameters, we define the wave-number
\mbox{$\klight\equiv|\Klight|=\omega/c$} and the wavelength
$\lambda\equiv2\pi/k$, where $c$ is the speed of light. We assume
that $E_0>0$ since the overall sign of the field can be incorporated
into the polarization $\ehat$. Equation (\ref{eq:E}) is appropriate
for the case of a light field with a Gaussian profile. The
discussion is limited to optical frequencies. In this regime,
$\Elight(\rvec,t)$ is uniform over atomic length scales. In order to
study the influence of the magnetic structure of the transition, we
also assume the presence of a uniform dc magnetic field \Bfield. The
quantization axis ($z$-axis) is chosen so that $\Bfield=B\,\zhat$
for $B>0$.}

The dynamics of an atom in the presence of the external magnetic and
electric fields described above is governed by the total Hamiltonian
\begin{equation}\label{eq:H(t)}
H= H_0 +H_1(t),
\end{equation}
where $H_0 \equiv \Hatomic+\HZeeman+\Hdc$  and $H_1(t) \equiv
\Hac(t)$ are the time-independent and time-dependent parts of $H$.
Here $\Hatomic$ is the atomic Hamiltonian, \mbox{$\HZeeman =
-\Mdipole\cdot\Bfield$} is the Zeeman Hamiltonian, \mbox{$\Hdc =
-\Edipole\cdot\Efield$} and \mbox{$\Hac(t) =
-\Edipole\cdot\Elight(\rvec,t)$} are the dc and ac Stark
Hamiltonians, and $\Mdipole$ and $\Edipole$ are the magnetic and
electric dipole moments of the atom, respectively. We assume that
\Bfield, \Efield, and $\Elight(\rvec,t)$ are sufficiently weak that
$\HZeeman$, $H_{\mrm{dc}}$, and $\Hac$ can be treated as successive
perturbations to $\Hatomic$.

Let $\ket{\ell}=\ket{\gamma J M}$ and $\EA(\ell)=\EA(\gamma J)$
represent the degenerate eigenstates of the atomic Hamiltonian
$\Hatomic$ and their corresponding energies, respectively.  Here $J$
is the total angular momentum quantum number,
\mbox{$M\in\{J,J-1,\hdots,-J\}$} is the magnetic quantum number
corresponding to the projection of the total angular momentum along
the $z$-axis, and $\gamma$ is a set of other quantum numbers.
\drdf{Then, to lowest order in the perturbing fields $\bfield$ and
$\efield$, the eigenstates of $H_0$ are}
\begin{equation}\label{eq:ellbar}
\ket{\overline{\ell}} = \ket{\ell} + \sum_{\ell'\neq\ell}
\ket{\ell'}\frac{\bra{\ell'}\Hdc|\ket{\ell}}{\EA(\ell)-\EA(\ell')},
\end{equation}
with corresponding energies
\begin{equation}\label{eq:E(ellbar)}
\mathcal{E}(\overline{\ell})= \EA(\ell)+\EZ(\ell)+\Edc(\ell).
\end{equation}
Here $\EZ(\ell)$ and $\Edc(\ell)$ represent the Zeeman and dc Stark
shifts, respectively. The Zeeman shift is given by \mbox{$\EZ(\ell)
= g_\ell\muB \bfield M$}, where $g_\ell$ is the Land\'e factor of
the state $\ket{\ell}$ and $\muB$ is the Bohr magneton. Throughout
this work, we assume that $B$ is sufficiently strong to completely
isolate the Zeeman sublevels of $\ket{\ell}$. The dc Stark shift is
given by $\Edc(\ell)=-(1/2)\adc_\ell\efield^2$, where
$\efield=|\Efield|$ is the magnitude of the dc electric field, and
$\adc_\ell$ is the dc polarizability of the atom in state
$\ket{\ell}$. To derive Eq.~(\ref{eq:ellbar}), we neglected mixing
of atomic eigenstates due to the magnetic field.

The atomic energy levels are also shifted by the ac Stark
shift~\cite{ref:Autler1955,ref:Angel1968,ref:Bonin1994,ref:Bonin1997},
which is induced by the dynamic field $\Elight(\rvec,t)$. We assume
that the frequency of the standing wave satisfies
$\omega\approx\omega_{ag}$, where $\omega_{ag}$ is the resonant
frequency of the electric-dipole (E1) transition from the perturbed
electronic ground state $\ket{\overline{g}}$ to a perturbed excited
state $\ket{\overline{a}}$. Thus it is appropriate to make a
two-level approximation that involves neglecting dynamic
interactions between states other than $\ket{\overline{g}}$ and
$\ket{\overline{a}}$. However, such an approximation can only
account for ac Stark shifts that arise due to mixing of the states
$\ket{\overline{g}}$ and $\ket{\overline{a}}$ with each other. To
address this situation, we modify the energy of the perturbed ground
state as follows:
\begin{equation}
\mathcal{E}(\overline{g})\rightarrow \mathcal{E}(\overline{g}) +
\Eac(\overline{g}),
\end{equation}
where
\begin{equation}\label{eq:StarkShift}
\Eac(\overline{g}) = -\frac{1}{2}\aac_{\overline{g}} E(\rvec)^2,
\end{equation}
is the ac Stark shift of $\ket{\overline{g}}$ due to mixing of
$\ket{\overline{g}}$ with states other than $\ket{\overline{a}}$,
and $\aac_{\overline{g}}$ is the corresponding ac polarizability of
the state $\ket{\overline{g}}$. An analogous modification is made to
the energy of the perturbed state $\ket{\overline{a}}$.

We further assume that the unperturbed ground and excited states
\mbox{$\ket{g}=\ket{\gamma_g J_g M_g}$} and
\mbox{$\ket{a}=\ket{\gamma_a J_a M_a}$} have the same parity. In
this case, the $\overline{g}\rightarrow \overline{a}$ transition is
induced by the dc electric field~\cite{ref:Bouchiat1975} and the
mixing of the states $\ket{\overline{g}}$ and $\ket{\overline{a}}$
with each other is characterized by the induced dipole matrix
element
\begin{equation}\label{eq:dag}
\bra{\overline{a}}\Edipole\cdot\ehat\ket{\overline{g}} \equiv \dind.
\end{equation}
We assume $\dind \geq0$ since any complex phase can be incorporated
into the states $\ket{\overline{a}}$ and $\ket{\overline{g}}$. Note
that the value of $\dind $ depends on the dc field $\Efield$ and the
light polarization $\ehat$. In particular, \mbox{$\dind\rightarrow
0$} as \mbox{$\Efield\rightarrow\mathbf{0}$}. Therefore, the dynamic
field does not cause mixing of the states $\ket{\overline{g}}$ and
$\ket{\overline{a}}$ in the absence of the electric field. \drdf{The
polarizability $\aac_{\overline{g}}$ in Eq.~(\ref{eq:StarkShift})
represents the ac polarizability of the unperturbed ground state
$\ket{g}$, provided the effects of the dc field on the ac
polarizability can be neglected.} In this case, $\aac_{\overline{g}}
= \aac_g$.

In general, ac polarizabilities depend on the polarization $\ehat$
and frequency $\omega$ of the external light field. The
polarizability of an arbitrary atomic state $\ket{\ell}$ can be
decomposed into three terms:
\begin{align}
\alpha_\ell&=\aac_0(\gamma J)+i\,\aac_1(\gamma J)\,\frac{M}{J}\,
(\ehat\times\ehat^{\ast})\cdot\zhat\nonumber\\&\quad +\aac_2(\gamma
J)\,\frac{3M^2-J(J+1)}{J(2J-1)}\, \frac{3|\ehat\cdot\zhat|^2-1}{2}.
\label{eq:aac}
\end{align}
The quantities $\aac_0(\gamma J)$, $\aac_1(\gamma J)$, and
$\aac_2(\gamma J)$ are referred to as the \emph{scalar},
\emph{vector}, and \emph{tensor} polarizabilities,
respectively~\cite{ref:Bonin1994}.  The scalar, vector, and tensor
polarizabilities are independent of the magnetic quantum number $M$
and the polarization $\ehat$, and hence are independent of the
choice of quantization axis and field geometry.  However, they
depend on the light frequency $\omega$, as described in
Appendix~\ref{app:A}.

% ==== FIGURE: Energy Level Diagram ====
\begin{figure}[tb]\center
\includegraphics[width=\columnwidth]{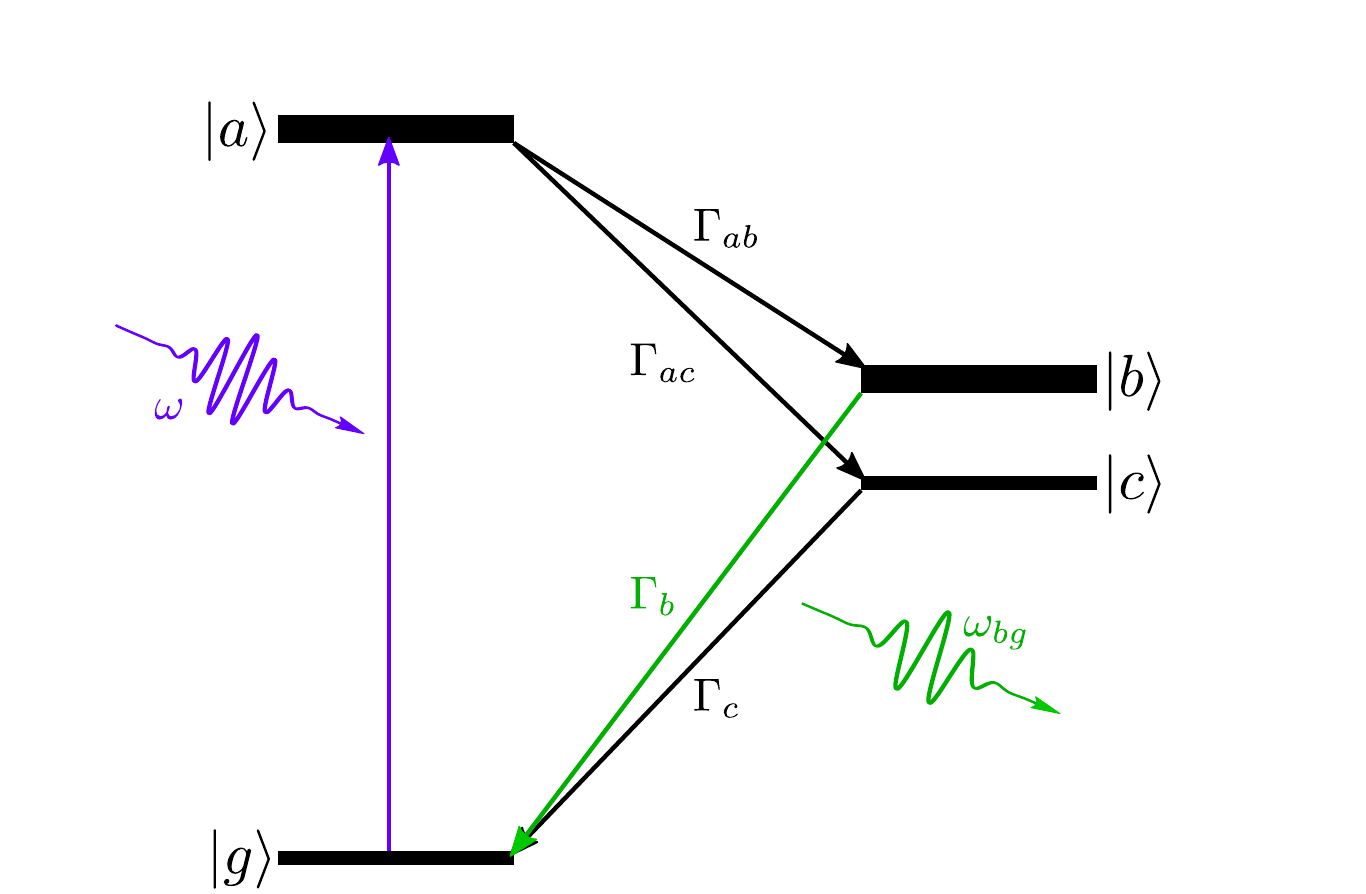}
\caption{\label{fig:EnergyLevelDiagram} \emph{Energy-level diagram.}
Shown are the energy eigenstates of an atom and the electronic
transitions relevant to the LSM.}
\end{figure}

Hereafter, we abandon the use of the overline to distinguish between
perturbed and unperturbed atomic states. Despite the lack of an
overline, quantum states should be interpreted as atomic states that
have been perturbed by the static electric and magnetic fields
\Efield and \Bfield, unless otherwise noted.

We consider a system in which atoms in the excited state undergo
spontaneous decay to the lower states $\ket{b}$ and $\ket{c}$ with
corresponding rates $\Gamma_{ab}$ and
$\Gamma_{ac}=\Gamma_a-\Gamma_{ab}$, where $\Gamma_a$ is the natural
linewidth of the state $\ket{a}$.  A schematic of the relevant
energy level structure is shown in
Fig.~\ref{fig:EnergyLevelDiagram}. As atoms decay from $\ket{b}$
down to $\ket{g}$, they emit fluorescent light of frequency
$\omega_{bg}$. The LSM involves both the simulation and measurement
of the spectral lineshape of the $g\rightarrow a$ transition. In
this context, the ``spectral lineshape" refers to the probability of
emission of fluorescent light of frequency $\omega_{bg}$ as a
function of laser frequency $\omega$. Although polarizabilities
$\aac_g$ and $\aac_a$ depend on $\omega$, we assume that they are
effectively constant for $\omega\approx\omega_{ag}$. The LSM can be
applied to any atomic system with the energy level structure shown
in Fig. \ref{fig:EnergyLevelDiagram}.

\section{Spectral lineshape}\label{sec:Lineshape}

% ==== FIGURE: Beam Orientation ====
\begin{figure}[tb]\center
\includegraphics[width=\columnwidth]{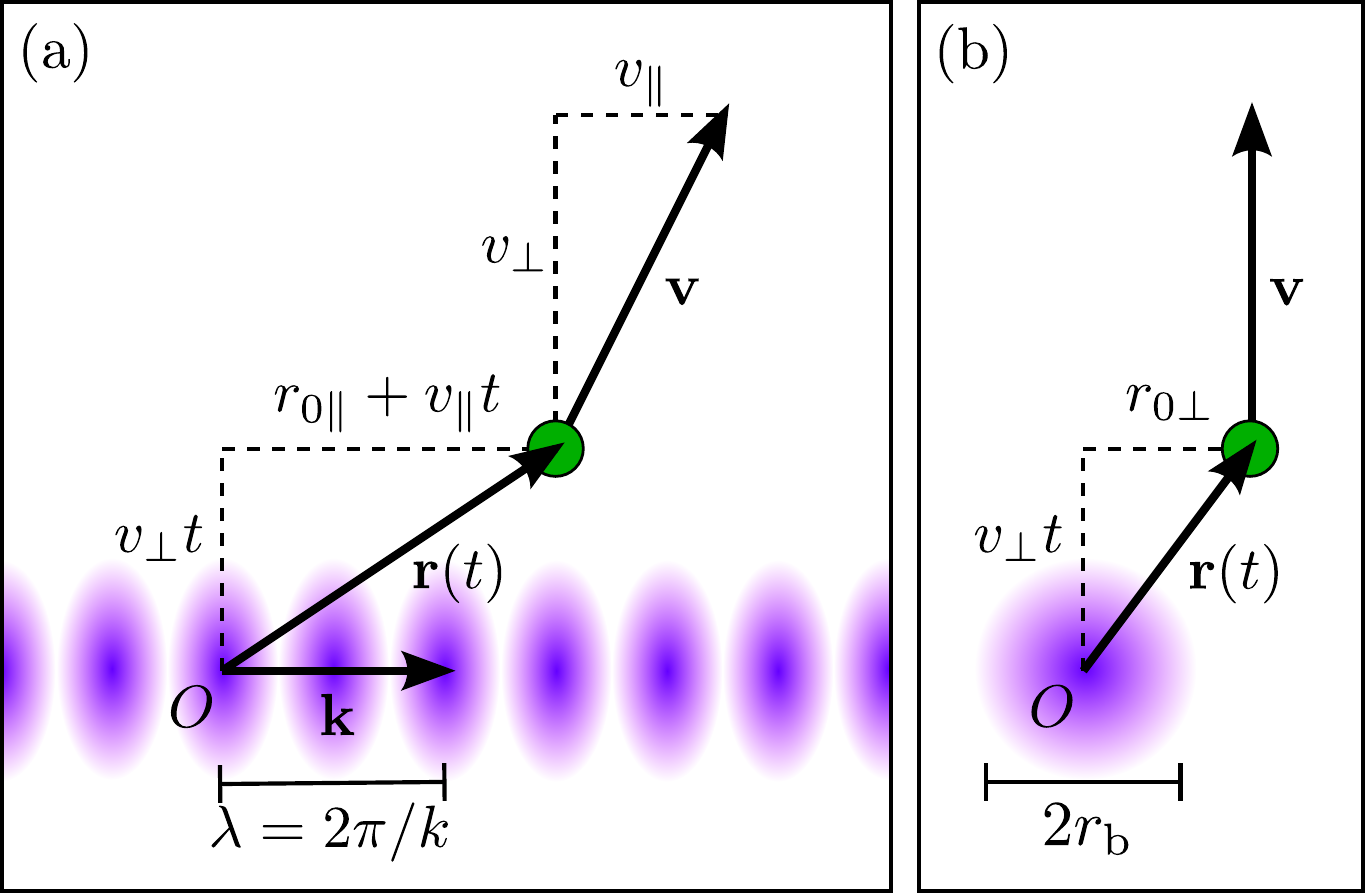}
\caption{\label{fig:Beam} \emph{Parameters of atomic trajectory.}
The atom's position $\mbf{r}(t)$ and velocity $\mbf{v}$ are shown
for some time $t>0$ and have been projected onto two planes: (a) the
plane formed by the vectors \Klight and $\mbf{v}$, where \Klight is
the wave-vector of the standing wave of light, and (b) the plane
normal to $\Klight$ and containing the origin $O$. A contour plot of
the spatial extent of the light intensity is shown for reference;
white and purple indicate minimal (zero) and maximal intensity,
respectively. Here $r_{0\parallel}$ and $r_{0\perp}$ correspond to
the components of $\rvec(t)$ that are parallel and perpendicular to
\Klight when $t=0$.}
\end{figure}

The ac Stark shifts cause the resonant frequency of the
$g\rightarrow a$ transition to shift as atoms travel through the
standing wave. As a result, the spectral lineshape depends heavily
on the details of the light field. Because the light field amplitude
\mbox{$E(\rvec)$} is not spatially uniform [see Eq. (\ref{eq:E}) and
the discussion thereafter], an atom with \drdf{coordinate}
$\rvec(t)$ will experience a time-dependent electric field in its
rest frame.

Assuming constant velocity, \drdf{the atom's position} is
$\mbf{r}(t)=\mbf{v}t+\mbf{r}_0$ where $\mbf{v}$ is the atomic
velocity and $\mbf{r}_0$ is the position of the atom when $t=0$. The
origin $O$ is chosen to be at one of the nodes of the standing wave.
A diagram of the geometry is shown in Fig.~\ref{fig:Beam}. The
time-dependent field experienced by the atom is given by
\begin{equation}\label{eq:E(t)}
\elight[\rvec(t)]= E_0\cos(\klight v_{\parallel}t+\klight
r_{0\parallel})\, e^{-(v_{\perp}^2t^2+r_{0\perp}^2)/\rbeam^2},
\end{equation}
where $\elight(\rvec)$ is given by Eq.~(\ref{eq:E}). Here
$v_{\parallel} = \hat{\Klight}\cdot\mbf{v}$ and
$v_{\perp}=|\mbf{v}-\hat{\Klight} \,v_{\parallel}|$ are the
components of the velocity that are perpendicular and parallel to
\Klight. Similarly, $r_{0\parallel}$ and $r_{0\perp}$ correspond to
the perpendicular and parallel components of the position $\rvec_0$.

\drdf{The total time dependence of the electric field in
Eq.~(\ref{eq:StandingWave}) is due to the fast oscillation of the
light at frequency $\omega$ and the slow modulation of the amplitude
with a frequency $kv_\parallel$. The amplitude modulation is
additionally characterized by a temporal Gaussian envelope with a
characteristic width $\rbeam/v_\perp$, which is the amount of time
an atom spends within the radius of the standing wave. We consider
non-relativistic atoms for which the conditions $\omega\gg k
v_\parallel$ and $\omega\gg v_\perp/\rbeam$ are valid, and the
optical oscillations are much faster than the modulation of the
amplitude $E[\rvec(t)]$. In this case, the ac Stark shift is
obtained by substituting Eq.~(\ref{eq:E(t)}) into
Eq.~(\ref{eq:StarkShift}).}

\subsection{Absorption profile}\label{sec:FM}
In order to gain a qualitative understanding of the physics, we make
the simplifying assumption
\begin{equation}\label{eq:E(t)approx}
E(t)\approx E_0\cos(\klight v_{\parallel}t).
\end{equation}
In this case, the Stark shifts of the ground and excited states lead
to the following shift of the resonant energy of the $g\rightarrow
a$ transition:
\begin{equation}\label{eq:shift}
\Eac(a)-\Eac(g)=-\frac{1}{2}\alpha_{ag}E_0^2 \cos^2(\klight
v_{\parallel}t),
\end{equation}
where
\begin{equation}
\alpha_{ag}\equiv\alpha_a-\alpha_g,
\end{equation}
is the difference of the polarizabilities of the ground and excited
states. From an atom's perspective, this is equivalent to a
polarizability-dependent frequency modulation of the two
counterpropagting light fields. Thus, the features of the
$g\rightarrow a$ lineshape can be understood by studying a related
system: stationary atoms with fixed energy levels in the presence of
two counter-propagating, frequency-modulated electric fields. In
this subsection, we turn our attention to such a system.

The frequency-modulated electric fields have instantaneous
frequencies $\omega_+$ and $\omega_-$ given by
\begin{equation}\label{eq:omegaeff}
\omega_\pm(t) = \omega \pm \klight v_{\parallel} +
\omegam\Am\cos(\omegam t) ,
\end{equation}
where
\begin{equation}\label{eq:omegam}
\omegam\equiv2\klight  v_{\parallel} \quad\mrm{and}\quad \Am \equiv
\frac{\aac_{ag}E_0^2}{8\hbar\klight  v_{\parallel}},
\end{equation}
are the modulation frequency and modulation index, respectively.
Equation (\ref{eq:omegaeff}) includes the term $\pm \klight
v_{\parallel}$ which accounts for the Doppler shifts of the
frequencies of the two counter-propagating waves. To derive
Eq.~(\ref{eq:omegaeff}), we used a trigonometric identity to write
$\cos^2(\klight  v_{\parallel}t)= (1/2)[1+\cos(2\klight
v_{\parallel}t)]$ and \drdf{we neglected the time-independent term
because it can be interpreted as an overall shift of the optical
frequency: $\omega\rightarrow\omega+\omegam\Am$.} The instantaneous
frequency $\omega_\pm(t)$ is characteristic of a light field with a
time-dependent phase~\cite{ref:Silver1992}. Such a field is given by
\begin{equation}\label{eq:Eeff1}
E_\pm = (1/2)E_0e^{i[(\omega\pm \klight v_\parallel) t+\Am
\sin(\omegam t)]},
\end{equation}
where the factor of 1/2 is included so that the total field, which
is the sum of two traveling waves, has an amplitude of $E_0$. The
effective field $E_{\pm}$ can be decomposed in the following way:
\begin{equation}\label{eq:Eeff}
E_{\pm} = \sum_{n=-\infty}^{\infty}(1/2)E_0 J_n(\Am)e^{i(\omega \pm
kv_{\parallel}+ n\omegam)t},
\end{equation}
where $J_n$ \drdf{are} Bessel \drdf{functions} of the first kind.
Thus the effective field consists of a \drdf{principal} field
($n=0$) which oscillates at a frequency $\omega\pm \klight
v_{\parallel}$, and infinitely many sidebands \mbox{$(n\neq 0)$}
which oscillate at frequencies \mbox{$\omega\pm \klight
v_{\parallel}+n\omegam$}. The amplitude of the electric field of
$n$th sideband is $(1/2)E_0J_{n}(\Am)$.

\drdf{The total electric field $E_{\mrm{tot}}$ seen by the atom is
the sum of the two counter-propagating light fields. To add the
fields, the summation index in Eq.~(\ref{eq:Eeff}) is changed from
$n$ to $n+1$ and the field $E_-$ is expressed as}
\begin{equation*}
E_{-} =\sum_{n=-\infty}^{\infty}(1/2)E_0 J_{n+1}(\Am)e^{i(\omega +
kv_{\parallel}+ n\omegam)t},
\end{equation*}
\drdf{where $-kv_\parallel+\omegam=+kv_\parallel$ because the
modulation frequency $\omegam$ is exactly twice the Doppler shift.
Hence the sidebands of the two counter-propagating waves overlap and
the total field is given by}
\begin{equation}\label{eq:Etot}
E_{\mrm{tot}} = E_{+} + E_{-}=\sum_{n=-\infty}^{\infty}\mathcal{E}_n
e^{i\omega_nt},
\end{equation}
where
\begin{equation}
\mathcal{E}_n \equiv \frac{E_0}{2}[J_{n+1}+J_{n}]
\quad\mrm{and}\quad \omega_n \equiv
\omega+kv_\parallel+2n\,kv_{\parallel}.
\end{equation}
\drdf{In particular, the first-order sidebands from one field
correspond with the carrier of the other~\cite{ref:Stalnaker2006},
resulting in an absorption profile with a polarizability-dependent
distortion. The ``absorption profile" is a plot of the transition
rate as a function of $\omega$.}

\begin{figure}[tb]\center
\includegraphics[width=\columnwidth]{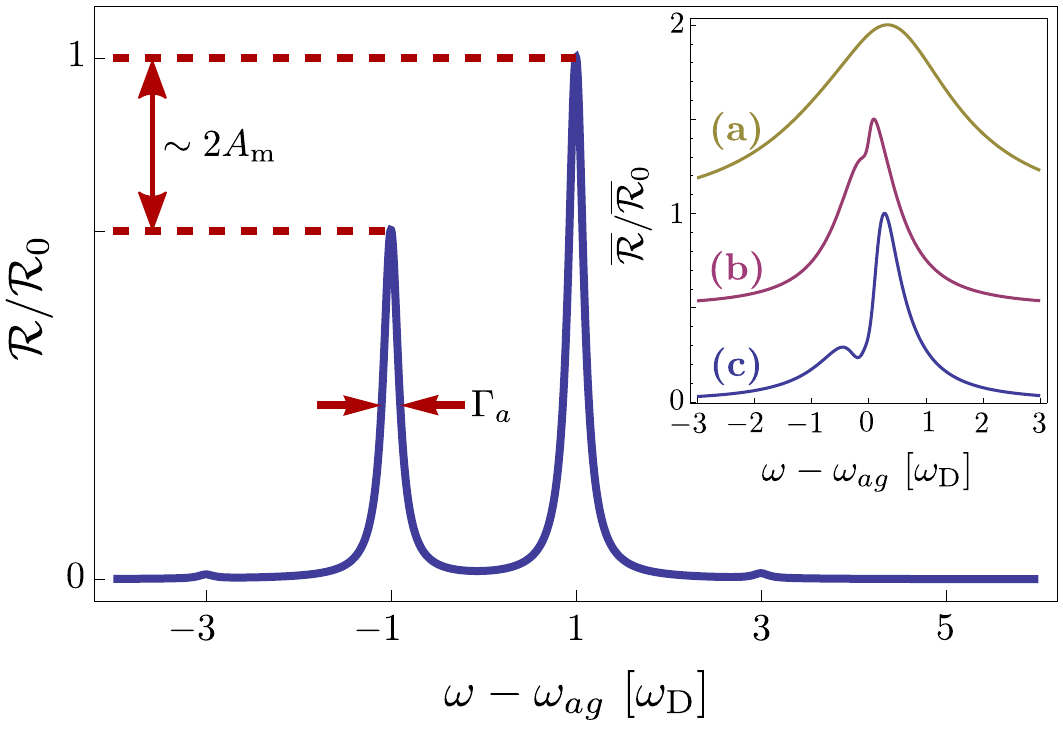}
\caption{\label{fig:ExcitationProbability} \emph{Absorption
profile.} Shown is a plot of the transition rate $\mathcal{R}$ as a
function of the laser frequency $\omega$ for the case of a single
atom with parallel speed $v_{\parallel}=\omegaD/k$, where $\omegaD$
is the Doppler broadening of the line. The transition rate is
normalized by $\mathcal{R}_0$ to have a maximum value of unity. Here
the modulation index satisfies \mbox{$A_{\mrm{m}}<0$}. For
\mbox{$A_{\mrm{m}}>0$}, the peak on the left is taller than the peak
on the right. In the insert, the average transition rate
$\overline{\mathcal{R}}$ is shown in three cases: (a) the condition
\mbox{$\Gamma_a/4\ll\omegaD$} is violated, (b) the condition
\mbox{$\hbar\omegaD\lesssim|\aac_{ag}|E_0^2$} is violated, and (c)
both conditions in Eq.~(\ref{eq:conditions}) are satisfied. To
generate these curves, we made use of the approximation
\mbox{$\klight\approx\klight_{ag}$} which is valid in the
near-resonant regime \mbox{($\omega\approx\omega_{ag}$)}.}
\end{figure}

The rate of the $g\rightarrow a$ transition is given by
\begin{equation}\label{eq:Rtot}
\mathcal{R} = \sum_{n=-\infty}^{\infty}
\frac{2\pi}{\hbar^2}\left|\dind
\mathcal{E}_n\right|^2\frac{\Gamma_a/2\pi}
{(\omega_n-\omega_{ag})^2+(\Gamma_a/2)^2}.
\end{equation}
Equation (\ref{eq:Rtot}) is valid in the weak excitation limit, that
is, when the excitation rate $\mathcal{R}$ is much smaller than all
other relevant rates. \drdf{To derive Eq.~(\ref{eq:Rtot}), we
neglected the interference of different harmonic components,
\emph{e.g.}, $\mathcal{E}_n e^{i\omega_nt}$ and $\mathcal{E}_{n'}
e^{i\omega_{n'}t}$. Such terms contribute small corrections to the
transition rate which do not affect the qualitative behavior of the
absorption profile.} A plot of the absorption profile is given in
Fig.~\ref{fig:ExcitationProbability}. The single-atom absorption
profile is clearly asymmetric about the atomic resonance
($\omega=\omega_{ag}$).

For an ensemble of atoms, the absorption profile is obtained by
averaging the transition rate (\ref{eq:Rtot}) over the velocity
distribution. The average rate is
\begin{equation}\label{eq:Rbar}
\overline{\mathcal{R}}\equiv \int_{-\infty}^{\infty}\!\!
\mathcal{R}(v_{\parallel})
\,\wcoll(v_{\parallel},\vC)\,dv_{\parallel},
\end{equation}
where $\wcoll(v_{\parallel},\vC)$ is the appropriate probability
distribution for the parallel velocity $v_{\parallel}$, and $\vC$ is
a characteristic speed in the $\hat{\Klight}$ direction. A plot of
the average transition rate is shown in the insert of
Fig.~\ref{fig:ExcitationProbability}.  For the insert,
$\wcoll(v_{\parallel},\vC)$ is taken to be a Lorentzian distribution
with a full-width at half the maximum value (FWHM) of $\vC=\omegaD/k
$, where $\omegaD$ is the overall Doppler broadening of the line.
\drdf{The resulting absorption profile is similar to experimentally
observed lineshapes in Yb (Section \ref{sec:App}).}

The absorption profile exhibits a polarizability-dependent feature:
a dip that separates the profile into two distinct peaks. The sign
of the polarizability $\alpha_{ag}$ determines whether the peak on
the left is larger or smaller than the peak on the right. The
following conditions need to be met in order for the distortion to
be observed:
\begin{equation}\label{eq:conditions}
\Gamma_{a}/4 \ll \omegaD \quad\mrm{and}\quad \hbar\omegaD \lesssim
|\alpha_{ag}|E_0^2.
\end{equation}
The former condition ensures that the sidebands of the FM waves can
be resolved. The latter ensures that the amplitude of the
first-order sidebands is not negligible compared to the amplitude of
the carrier, that is, $A_{\mrm{m}}\neq 0$ for
$v_{\parallel}\approx\vC$. If either $\omegaD\lesssim\Gamma_a/4$ or
$\hbar\omegaD\gg |\aac_{ag}|E_0^2$, then the asymmetric distortion
will be suppressed, as can be seen in the insert of
Fig.~\ref{fig:ExcitationProbability}. In this case, the methods
described here cannot be used to measure the polarizability
$\alpha_{ag}$. However, the LSM can still be used to measure
$\alpha_{ag}$ in the absence of the distortion by comparing
displacements of the central peak of the lineshape.

\drdf{By omitting the Gaussian envelope
$\exp(-v_\perp^2t^2/\rbeam^2)$ in Eq.~(\ref{eq:E(t)approx}), we
neglect effects of the atom's finite transit time, such as
broadening of the spectral line~\cite{ref:Demtroder2003}.
Nonetheless, the transit time must satisfy the following
restrictions:}
\begin{equation}\label{eq:conditions2}
v_\perp/\rbeam \lesssim \dind \elight_0/\hbar\quad\mrm{and}\quad
v_\perp/\rbeam \ll \omegaD,
\end{equation}
\drdf{where $\dind \elight_0/\hbar$ is the Rabi frequency of the
$g\rightarrow a$ transition. The former condition represents a
system in which atoms have enough time to undergo excitation to the
upper state $\ket{a}$, as will be discussed in
Section~\ref{sec:Model}. The latter condition ensures that the
Doppler broadening is sufficiently large that most atoms travel
through many nodes and antinodes of the standing wave during their
transit.}

Although the absorption profile provides a satisfactory illustration
of the physics, it cannot be used to measure the ac
polarizabilities. The transition rate presented in
Eq.~(\ref{eq:Rtot}) is valid only in the weak excitation limit and
therefore cannot account for saturation effects. Moreover,
Eq.~(\ref{eq:Rtot}) does not properly take into account interference
of different probability amplitudes, \drdf{nor does it include
finite transit-time effects.} A more complete picture is required to
generate a theoretical lineshape that can be fitted to experimental
data. Such a picture is achieved by the following model.

% ====================================
% MODEL
% =====================================
\subsection{Fluorescence probability}\label{sec:Model}
Hereafter, we return our attention to the original system: moving
atoms illuminated by light with a fixed frequency $\omega$. The
spectral lineshape of the $g\rightarrow a$ transition is modeled by
computing the probability of emission of fluorescent light of
frequency $\omega_{bg}$ as a function of laser frequency $\omega$.
The computation involves three steps. First, the time-dependent
population $\rho_{bb}$ of the state $\ket{b}$ is computed by
numerically solving the optical Bloch equations (OBE) for the case
of atoms traveling through a non-uniform field (see
Fig.~\ref{fig:Beam}). Second, the probability of fluorescence
$\mathcal{F}$ is determined by integrating the time-dependent decay
rate $\Gamma_b\rho_{bb}$ with respect to time. Finally, the average
fluorescence probability $\overline{\mathcal{F}}$ is computed by
taking a weighted average of $\mathcal{F}$ with respect to the
atomic velocity $\mbf{v}$ and the offset $\mbf{r}_0$.

Although the LSM can be used in conjunction with any atomic source,
our model makes use of distributions that are appropriate for a
collimated beam of thermal atoms traveling in a direction that is
orthogonal to the standing wave. In this case, $v_{\perp}$
represents the component of the atom's velocity along the atomic
beam, and $v_{\parallel}/v_{\perp}$ represents the angular spread of
the beam. The corresponding velocity distribution is
\begin{equation}\label{eq:wv}
w(\mbf{v}) = \woven(v_{\perp},\vT)\, \wcoll(v_{\parallel},\vC),
\end{equation}
where
\begin{equation}\label{eq:woven}
\woven(v_{\perp},\vT) = 2(v_{\perp}^3/\vT^4)e^{-(v_{\perp}/\vT)^2},
\end{equation}
is the distribution of velocities appropriate for thermal atoms
escaping from a hole, and $\wcoll(v_{\parallel},\vC)$ is the
velocity distribution appropriate for a collimated atomic beam. Here
$\vT=\sqrt{2\kB T/M}$ is the thermal speed of the atom, $T$ is the
temperature of the oven, $M$ is the atom's mass, $\kB$ is
Boltzmann's constant, and $\vC$ is the characteristic speed
determined by the atomic-beam collimator.  To model the effects of a
vane collimator, we approximate the spread of parallel velocities by
a Lorentzian distribution with a FWHM of $\vC=\omegaD/k $, where
$\omegaD$ is the overall Doppler broadening of the line.

In the following model, we use dimensionless parameters.
Dimensionless parameters ease computation, and potentially
facilitate the application of the model to several different atomic
systems. Throughout, we make the approximation
$\klight\approx\klight_{ag}$ which is valid in the near-resonant
regime ($\omega\approx\omega_{ag}$).

Time \drdf{is} measured in units $1/\Gamma_a$. We define the
dimensionless time $\tau\equiv\Gamma_a\,t$ and decay rates
\mbox{$G_b \equiv \Gamma_b/\Gamma_a$}, \mbox{$G_c \equiv
\Gamma_c/\Gamma_a$}, \mbox{$G_{ab} = \Gamma_{ab}/\Gamma_a$}, and
\mbox{$G_{ac} = 1-G_{ab}$}. We further define the dimensionless
perpendicular and parallel velocities \mbox{$u_{\perp} \equiv
v_{\perp}/\vT$} and \mbox{$u_{\parallel} \equiv v_{\parallel}/\vC$},
and the dimensionless perpendicular and parallel offsets
\drdf{\mbox{$\xi \equiv r_{0\perp}/\rbeam$} and \mbox{$\varphi
\equiv k r_{0\parallel}$}}, respectively.

Consistent with the discussion in Section~\ref{sec:FM}, we introduce
the following dimensionless parameters: the saturation parameter
$S_0$, \drdf{characteristic modulation index $S_1$, and Doppler
parameter $S_2$,} defined by
\begin{align}
S_0 &\equiv \left[\dind E_0/(\hbar\Gamma_a)\right]^2,\\
S_1 &\equiv \alpha_{ag}E_0^2/(8\hbar\omegaD),
\end{align}
and
\begin{equation}
S_2 \equiv 2\omegaD/\Gamma_a,
\end{equation}
respectively. We define an additional parameter $S_3$ by
\begin{equation}
S_3 \equiv \vT/(\rbeam\Gamma_a).
\end{equation}
\drdf{Note that $\rbeam/v_{\perp}$ is the time that an atom spends
within the radius of the light field} and hence $1/(S_3\,u_{\perp})$
represents the dimensionless transit time.

\begin{figure}[t]\center
\includegraphics[width=\columnwidth]{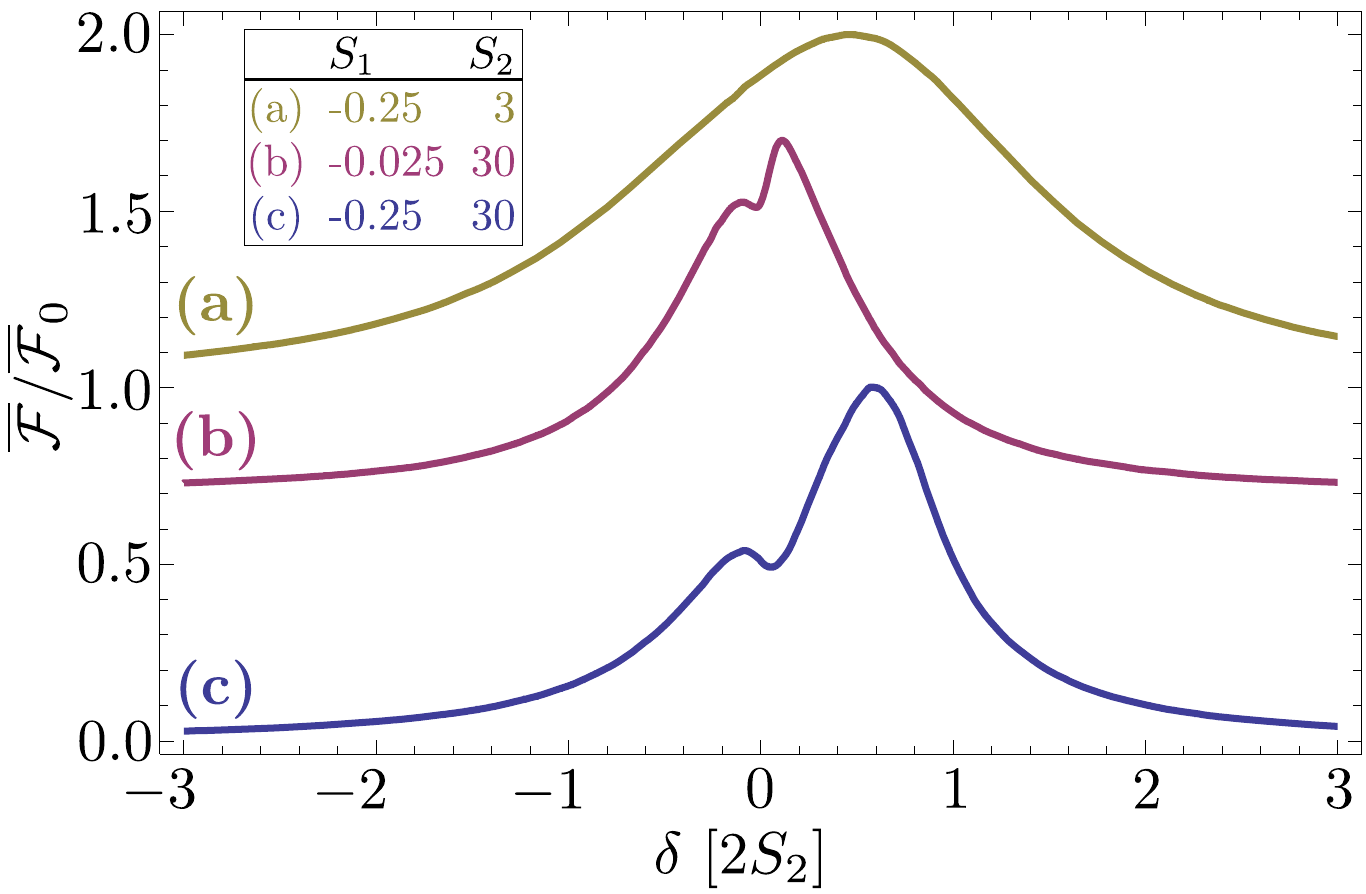}
\caption{\label{fig:lineshapes} \emph{Results of simulation.} The
simulated lineshape $\overline{\mathcal{F}}(\delta)$ is shown for
two different values each of the modulation index $S_1$ and the
Doppler parameter $S_2$, as indicated by the insert: (a) the
condition $S_2\gg 1/2$ is violated, (b) the condition $|S_1|\gtrsim
1/8$ is violated, and (c) both conditions are satisfied. For each
curve, the saturation parameter is $S_0=1.0$. The lineshapes are
normalized by the maximum peak height $\overline{\mathcal{F}}_0$. To
ease comparison, the middle and top curves are shifted vertically by
0.7 and 1.0, respectively. }
\end{figure}

In terms of the dimensionless parameters, the conditions presented
in expressions~(\ref{eq:conditions}) reduce to \mbox{$S_2\gg 1/2$}
and \mbox{$|S_1|\gtrsim 1/8$}. When either of these conditions is
violated, the characteristic dip in the lineshape is suppressed, as
can be seen in Fig.~\ref{fig:lineshapes}. Likewise,
conditions~(\ref{eq:conditions2}) reduce to $S_3\lesssim\sqrt{S_0}$
and $S_3\ll (1/2)S_2$. Whereas the absorption profile discussed in
Section~\ref{sec:FM} was valid only in the weak excitation limit
$(S_0\ll 1)$, the model of the fluorescence can accommodate large
saturation parameters.

\drdf{Let $\rho_{nm}$ be the elements of the density matrix in the
atom's rest frame for states $n,m\in\{g,a,b,c\}$. We assume that the
rotating wave approximation holds and dynamic interactions between
states other than $\ket{g}$ and $\ket{a}$ can be neglected. In this
case, the dimensionless optical Bloch equations (OBE) for the
configuration shown in Fig. \ref{fig:EnergyLevelDiagram}
are~\cite{ref:Loudon2000}}
\begin{subequations}\label{eq:sigma}
\begin{align}
\dot{\rho}_{aa} &= -\frac{i}{2}\Gammap(\rho_{ag}-\rho_{ga})-
\rho_{aa},\\
\dot{\rho}_{ag} &= +\frac{i}{2}\Gammap(\rho_{gg}-\rho_{aa})-
\frac{1}{2}(1-2i\Delta)\rho_{ag},\\
\dot{\rho}_{bb} &=-G_{b}\rho_{bb}+G_{ab}\rho_{aa},\\
\dot{\rho}_{cc} &=-G_{c}\rho_{cc}+G_{ac}\rho_{aa},
\end{align}
\end{subequations}
where \mbox{$\dot{\rho}_{nm} = d(\rho_{nm})/d\tau$}.  The remaining
density matrix elements $\rho_{gg}$ and $\rho_{ga}$ are determined
from \mbox{$\sum_n\rho_{nn} = 1$} and
\mbox{$\rho_{ga}=\rho_{ag}^{\ast}$}. Here
\begin{equation}
\Gammap \equiv
\frac{1}{\Gamma_a}\left[\frac{\bra{a}\Edipole\cdot\ehat\ket{g}\elight(t)}{
\hbar}\right] = \sqrt{S_0}\,f(\boldsymbol{\xi},\mbf{u}),
\label{eq:Gamma_pump}
\end{equation}
is the Rabi frequency,
\begin{equation}\label{eq:Delta}
\Delta \equiv
\frac{1}{\Gamma_a}\left[\omega-\frac{\Energy(a)-\Energy(g)}{\hbar}\right]=
\delta+ 2S_1S_2\,f(\boldsymbol{\xi},\mbf{u})^2,
\end{equation}
is the detuning of the laser light from the resonance, \mbox{$\delta
\equiv (\omega-\omega_{ag})/\Gamma_a$}, and the function
$f(\boldsymbol{\xi},\mbf{u})$ is defined by
\begin{equation}
f(\boldsymbol{\xi},\mbf{u}) \equiv \cos[(S_2/2)u_\parallel
\tau+\varphi ]e^{-[(S_3u_\perp\tau)^2+\xi^2]},
\end{equation}
where $\boldsymbol{\xi}=(\varphi ,\xi)$ and
$\mbf{u}=(u_{\parallel},u_{\perp})$. \drdf{We further assume that
all atoms initially occupy the ground state.}

The probability that an atom will emit fluorescent light of
frequency $\omega_{bg}$ in a time interval $[\ti,\tf]$ is given by
\begin{equation}\label{eq:Flo}
\mathcal{F}= \int_{\ti}^{\tf}G_b\,\rho_{bb}(\tau)\,d\tau,
\end{equation}
where $G_b\,\rho_{bb}(\tau)$ is the time-dependent rate of the
$b\rightarrow g$ decay.  The choice of integration interval
$[\ti,\tf]$ depends on both the transit time and the characteristic
time of the fluorescent decay after the atoms leave the light field.
In the case where the $a\rightarrow b\rightarrow g$ decay time is
shorter than the transit time, it is appropriate to define the
integration limits by \mbox{$-\ti=\tf=3/(S_3\,u_{\perp})$}.  The
factor of 3 ensures that the atom is ``far" from the standing wave
at the integration limits. To model a system with slower decays, the
integration limit $\tf$ must be extended.

The fluorescence probability depends on the parameters of the atomic
trajectory, that is,
$\mathcal{F}=\mathcal{F}(\boldsymbol{\xi},\mbf{u})$. We define the
average probability of fluorescence by
\begin{equation}\label{eq:Flobar}
\overline{\mathcal{F}}=
\iint\mathcal{F}(\boldsymbol{\xi},\mbf{u})\,w(\boldsymbol{\xi},\mbf{u})
\,d\boldsymbol{\xi}\,d\mbf{u},
\end{equation}
where $w(\boldsymbol{\xi},\mbf{u})$ is the probability distribution
associated with the atom's initial position and velocity.  In our
model, we assume $w(\boldsymbol{\xi},\mbf{u}) =
w_1(\boldsymbol{\xi})\,w_2(\mbf{u})$, where the distribution $w_1$
is a uniform distribution over the intervals $\xi\in[-3,3]$ and
$\varphi \in[-\pi,\pi]$. The finite integration limits are justified
by the following properties of the system: First, the amplitude of
the standing wave drops to less than 0.01\% of its maximum value
when $|\xi|>3$. Therefore, atoms will only pass through the light if
$|\xi|\lesssim3$. Second, $\varphi $ constitutes a phase shift of
the electric field which is unique only for $\varphi \in[-\pi,\pi]$.
Consistent with Eqs.~(\ref{eq:wv}) and (\ref{eq:woven}), the
velocity distribution satisfies
$w_2(\mbf{u})=\woven(u_{\perp},1)\wcoll(u_{\parallel},1)$, where
$\woven(u_{\perp},1)$ is the velocity distribution for atoms
escaping from a hole with unit thermal speed, and
$\wcoll(u_{\parallel},1)$ is a Lorentzian distribution with FWHM of
1.

The fluorescence probability is a function of the detuning $\delta$
from the atomic resonance, that is,
\begin{equation}
\overline{\mathcal{F}}  =
\overline{\mathcal{F}}(\delta;\mbf{S},\mbf{G}),
\end{equation}
where $\mbf{S}\equiv(S_0,S_1,S_2,S_3)$ and
$\mbf{G}\equiv(G_{ab},G_b,G_c)$ are parameters.  We refer to a plot
of $\overline{\mathcal{F}}(\delta;\mbf{S},\mbf{G})$ as a function of
$\delta$ as the ``simulated lineshape" of the $g\rightarrow a$
transition.  Three such plots are shown in
Fig.~\ref{fig:lineshapes}. In Fig.~\ref{fig:lineshapes}, curves~(a)
and~(b) demonstrate the suppression of the polarizability-dependent
distortion when either the Doppler parameter $S_2$ or the modulation
index $S_1$ is too small. The distortion is most pronounced in
curve~(c), for which $S_2\gg1/2$ and $|S_1|\gtrsim 1/8$. The
simulated lineshapes in Fig.~\ref{fig:lineshapes} are qualitatively
similar to the absorption profiles in the insert of
Fig.~\ref{fig:ExcitationProbability}, as expected.

\subsection{Numerical procedure}\label{sec:Num}
\drdf{The numerical procedures described here are valid for a
variety of atomic species. However, the simulations were performed
with parameter values appropriate for the Yb system described in
Section~\ref{sec:App}.}

We used a stiffly stable Rosenback
method~\cite{ref:NumericalRecipes} to numerically solve a system of
equations related to Eqs.~(\ref{eq:sigma}) and (\ref{eq:Flo}), with
\mbox{$-\ti=\tf=3/(S_3\,u_{\perp})$}. This system of equations is
described in Appendix~\ref{app:B}. The Rosenback method involves two
tolerances--denoted \texttt{atol} and \texttt{rtol} in
Ref.~\cite{ref:NumericalRecipes}--which were both set to $10^{-5}$.
The multi-dimensional integral in Eq.~(\ref{eq:Flobar}) was computed
using an adaptive Monte Carlo routine~\cite{ref:NumericalRecipes}.
In our implementation, the integration routine involves $10^5$
evaluations of the integrand. For various values of $\mbf{S}$, the
average estimated error was less than 1\% of the value of the
integral.

For computational purposes, we restricted the integration to the
following finite domain: $\xi\in[-3,3]$, $\varphi \in[-\pi,\pi]$,
$u_{\perp}\in[0,3]$, and $u_{\parallel}\in[-6,6]$. The subdomains
for $\xi$ and $\varphi $ were discussed after Eq. (\ref{eq:Flobar}).
The finite integration subdomains for $u_\parallel$ and $u_\perp$
are justified as follows:  Atoms with a large parallel speed
$|u_{\parallel}|$ experience a Doppler shift that is much larger
than the characteristic Doppler broadening of the spectral line.
Such atoms only contribute to the wings of the lineshape, where
$|\delta|$ is large and the probability of fluorescence is very
small. Moreover, for a Lorentzian velocity distribution with unit
FWHM, $|u_{\parallel}|\leq 6$ for about 95\% of atoms.

On the other hand, atoms with a perpendicular speed that satisfies
\mbox{$u_\perp\gg\sqrt{S_0}/S_3$} are moving so fast that the
transit time is much smaller than the inverse Rabi frequency. Such
atoms do not spend enough time in the light field for the
$g\rightarrow a$ transition to be realized. Since most atoms travel
at or near the thermal speed $u_\perp = 1$, the condition
$\sqrt{S_0}/S_3\gtrsim 1$ represents a system in which most atoms
have enough time to interact with the light.  We assume that atoms
with speed $u_\perp>3$ do not contribute significantly to the
lineshape.  Note that $u_\perp>3$ for only about 0.1\% of atoms.
\drdf{Finally, we ignore counterflow of atoms in the atomic beam by
requiring $u_\perp\geq0$.}

For fixed $\mbf{G}=(0.35,0.45,0)$ and $S_3=0.53$, we computed the
average fluorescence
$\overline{\mathcal{F}}(\delta;\mbf{S},\mbf{G})$ for 100 discrete
values of \mbox{$\delta\in[\delta_-,\delta_+]$}, and various
discrete values of $S_0\in[0.1,10]$, $S_1\in[0.01,1]$, and
$S_2\in[1,100]$. Here \mbox{$\delta_{\pm}\equiv
2S_1S_2\pm3[(1/2)S_2+G_b+1]$}. The results were interpolated using
cubic splines to approximate the continuous function
$\overline{\mathcal{F}}(\delta;\mbf{S},\mbf{G})$.  Three curves
which are typical of those produced by this procedure are presented
in Fig.~\ref{fig:lineshapes}.

The LSM involves fitting the simulated curve
\mbox{$\overline{\mathcal{F}}(\delta;\mbf{S},\mbf{G})$} to the
observed lineshape to determine best-fit values of $S_0$, $S_1$, and
$S_2$.  From the best-fit values, the following three quantities can
be calculated: the polarizability difference $\alpha_{ag}$, the
circulating power $P$ of the standing wave\footnotemark, and the
Doppler broadening $\omegaD$ of the $g\rightarrow a$ transition
which are given by
\begin{align}
\aac_{ag} &= \left[\frac{4\,\dind ^2}{\hbar\,\Gamma_a}\right]
\times S_1S_2/S_0,\label{eq:aacu}\\
P &= \left[\frac{1}{16}\pi\,\rbeam^2\,c\,\epsilon_0
\frac{\hbar^2\,\Gamma_a^2}{\dind ^2}\right]\times S_0\label{eq:Pu},
\end{align}
and
\begin{equation}
\omegaD = (\Gamma_a/2)\times S_2\label{eq:wu},
\end{equation}
respectively.  Here $\epsilon_0$ is the permittivity of free space.
The present implementation of the LSM uses \emph{Mathematica}'s
nonlinear regression routine to determine the best-fit values of
$S_0$, $S_1$, and $S_2$. {Alternatively, Eq.~(\ref{eq:Pu}) can be
solved for $\dind$ in terms of $P$ and $S_0$. Thus the LSM can also
be used to measure the induced dipole moment when the power is
known, as was done in the previous application of the
LSM~\cite{ref:Stalnaker2006}.}

The treatment of systematic errors and statistical uncertainties is
straightforward. In Eq.~(\ref{eq:aacu}), for instance, the
uncertainty in the dimensionful quantity $\dind ^2/(\hbar\Gamma_a)$
is due solely to systematic effects, whereas the uncertainty in the
term $S_1S_2/S_0$ arises from statistical uncertainties in both the
observed signal and the fitting algorithm. The total uncertainty of
the quantity $\alpha_{ag}$ is obtained by adding these independent
uncertainties in quadrature.  {If the signal-to-noise ratio (SNR) of
the observed lineshape is sufficiently high, then the error of the
measurement of $\aac_{ag}$ will be dominated by the uncertainties of
the known quantities $\dind$ and $\Gamma_a$.}

\footnotetext{The standing wave is formed by two counter-propagating
waves of light. The circulating power of the standing wave is
defined as the average power of a single traveling wave.  The
electric field of the wave propagating in the $\pm\hat{\Klight}$
direction is given by
\mbox{$E_1(\rvec,t)=(1/2)E_0\exp[-(r_{\perp}/\rbeam)^2]\cos(kr_{\parallel}\pm\omega
t)$}. The corresponding time-averaged intensity in SI units is
\mbox{$I(r_{\perp})=(1/8)c\epsilon_0E_0^2\exp[-2(r_{\perp}/\rbeam)^2]$},
where the time average is taken over a single period of oscillation.
Therefore, the circulating power of the standing wave is given by
\mbox{$P\equiv \int_{0}^{\infty}I(r_{\perp})\,2\pi
r_{\perp}\,dr_{\perp}=(1/16)\pi\rbeam^2\,c\epsilon_0 E_0^2$}.}

% =====================================
% APPLICATION
% =====================================
\section{Application to Ytterbium}\label{sec:App}
\begin{figure}[tb]\center
\includegraphics[width=\columnwidth]{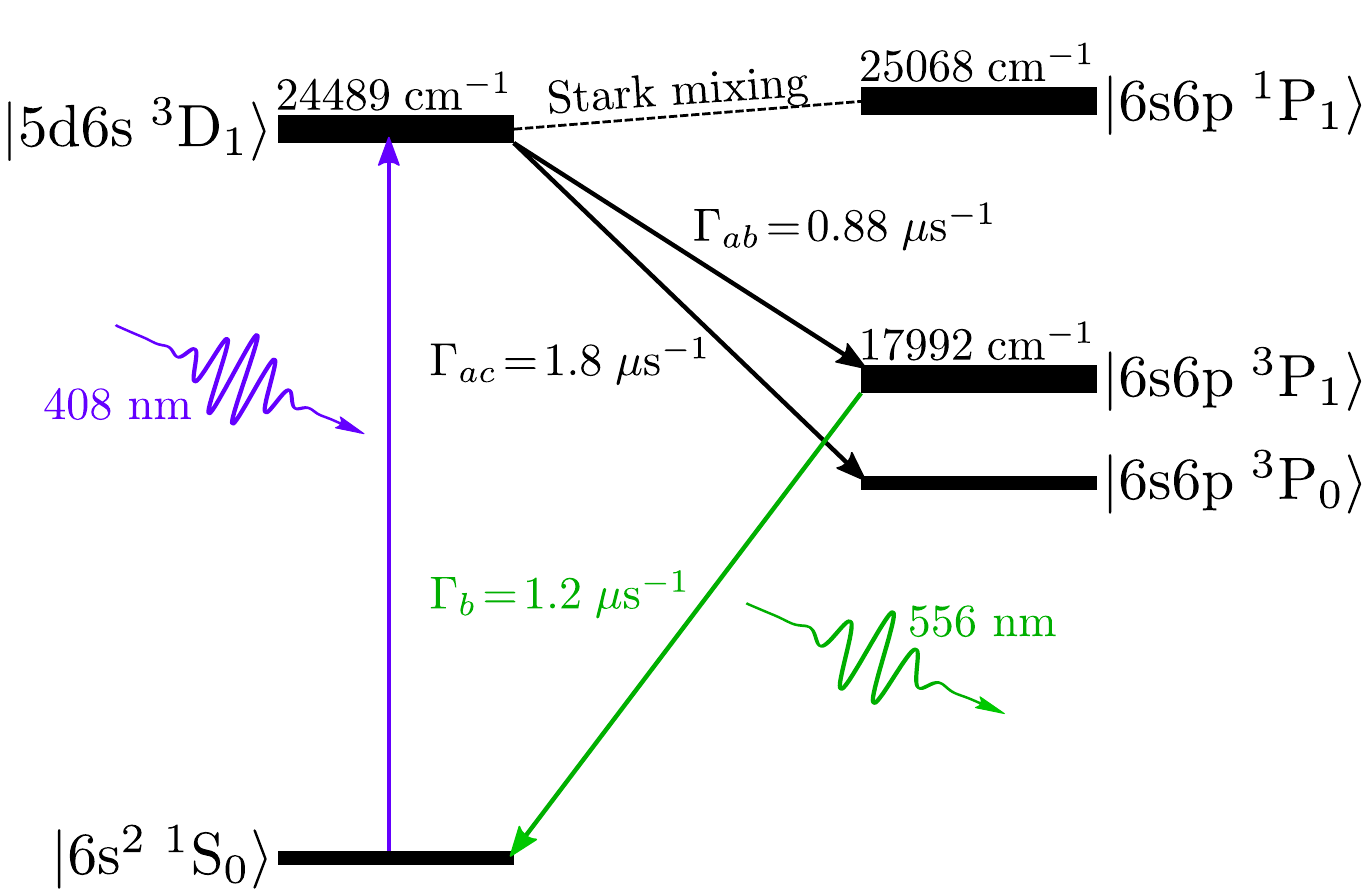}
\caption{\label{fig:YbEnergy} \emph{Ytterbium energy levels.} Shown
are the low-lying energy eigenstates of Yb and the electronic
transitions relevant to the polarizability experiment.}
\end{figure}

The electronic structure of Yb is shown in Fig.~\ref{fig:YbEnergy}.
The low-lying energy eigenstates of Yb match the structure shown in
Fig.~\ref{fig:EnergyLevelDiagram} under the following mapping:
\mbox{$\ket{g}=\ket{\SSzero}$}, \mbox{$\ket{a}=\ket{\TDone}$},
\mbox{$\ket{b}=\ket{\TPone}$}, and \mbox{$\ket{c}=\ket{\TPzero}$}.
Therefore, the LSM can be used to measure the difference in ac
polarizabilities of the upper state $\ket{\TDone}$ and the ground
state $\ket{\SSzero}$ at 408 nm by analyzing the lineshape of the
\mbox{408-nm} $\SSzero\rightarrow\TDone$ transition. In this case,
the lineshape is measured by observing the 556-nm fluorescence of
the $\TPone\rightarrow\SSzero$ decay. The intermediate state
$\ket{\TPzero}$ is metastable and hence $\Gamma_c\approx0$.

There is an additional decay of $\ket{\TDone}$ to the metastable
state $\ket{\TPtwo}$, which is not shown in
Fig.~\ref{fig:EnergyLevelDiagram}. The state $\ket{c}$ can represent
multiple metastable states, including both $\ket{\TPone}$ and
$\ket{\TPtwo}$. In this interpretation, $\Gamma_{ac}$ is the rate of
decay of $\ket{\TDone}$ to all metastable states.

 % ==== FIGURE: Yb Energy Level Diagram ====

The highly forbidden $\SSzero\rightarrow\TDone$ transition is
induced by the Stark mixing technique.  This technique involves the
application of a static, uniform electric field \Efield which mixes
the upper state $\ket{\TDone}$ with opposite-parity states,
predominantly the $\ket{\SPone}$ state. The effective dipole moment
$\dind $ associated with the Stark-induced transition is given
by~\cite{ref:Stalnaker2006}
\begin{equation}\label{eq:dagCag}
\dind =\beta_{ag}|\kq{(\Efield\times\ehat)}{1}{-M_a}|,
\end{equation}
where \mbox{$\beta_{ag}= 2.18(10)\times10^{-8}\;e\cdot
a_0\cdot(\mrm{V}/\mrm{cm})^{-1}$} is the vector transition
polarizability of the $\SSzero\rightarrow\TDone$ transition, $e$ is
the elementary charge, $a_0$ is the Bohr radius, and
$\kq{(\mbf{E}_{\mrm{dc}}\times\ehat)}{1}{}$ is a spherical tensor of
rank one\footnotemark. Here $M_a$ is the magnetic quantum number of
the $\ket{\TDone}$ state.

\footnotetext{Let $\kq{A}{1}{}$ be the rank-one spherical tensor
associated with the Cartesian vector $\mbf{A}$. Then the components
of $\kq{A}{1}{}$ are given by $\kq{A}{1}{0} = A_z$ and
$\kq{A}{1}{\pm1}=\mp(A_x\pm i A_y)/\sqrt{2}$.}

Because the angular momentum of the ground state $\ket{\SSzero}$ is
$J_g=0$, only the scalar term in Eq. (\ref{eq:aac}) contributes to
the polarizability of $\ket{\SSzero}$. That is,
\mbox{$\alpha_g=\alpha_0(\SSzero)$} and so $\aac_g$ is independent
of the geometry of the applied fields.  Hence the dependence of
$\alpha_{ag}$ on the field geometry is due entirely to the vector
and tensor polarizabilities $\alpha_1(\TDone)$ and
$\alpha_2(\TDone)$ of the excited state $\ket{\TDone}$. In this
case, the LSM is sensitive to the difference of the scalar
polarizabilities of the ground and excited states. However, the
vector and tensor polarizabilities of the excited state can be
measured unambiguously by varying the polarization of the standing
wave.

A recent calculation~\cite{ref:DereviankoPC} of the polarizability
of the ground state $\ket{\SSzero}$ at 408 nm yielded\footnotemark
\begin{equation}\label{eq:alpha0_estimate}
\aac_0(\SSzero) = 0.218(11)\;\mrm{Hz}\cdot\mrm{(V/cm)^{-2}}.
\end{equation}
Calculations of the polarizability $\ket{\TDone}$ at 408 nm are
complicated by the potential existence of odd-parity eigenstates
with energy close to twice the energy of a 408-nm photon.  Such
states could lead to a resonantly enhanced polarizability of the
$\TDone$ state.  The energy spectrum in this region (which is below
the ionization limit) is very dense due to the excitation of $4f$
orbitals.  The knowledge of the energy spectrum is far from complete
in this region.  This provides one of the motivations for
determining the polarizabilities experimentally.

\footnotetext{Equation~(\ref{eq:StarkShift}) implies that
$\alpha_{ag}$ has units of \emph{energy} per squared electric field.
However, in this work $\alpha_{ag}$ is normalized by $2\pi\hbar$ and
presented in units of \emph{frequency} per squared electric field.
For a more thorough discussion of unit conventions, see
Ref.~\cite{ref:Mitroy2010}.}

The first implementation of the LSM~\cite{ref:Stalnaker2006} was
used to measure the quantity
\begin{equation}\label{eq:aI}
\aac_{ag}^{\mrm{I}} =-0.312(34)\;\mrm{Hz}\cdot\mrm{(V/cm)^{-2}},
\end{equation}
were $\aac_{ag}^{\mrm{I}} = \aac_2(\TDone)+ \aac_0(\TDone)
-\aac_0(\SSzero)$.  Here the superscript ``I" is introduced to
distinguish this measurement from the results of the present work.
To determine the tensor contribution $\aac_2(\TDone)$ unambiguously
requires a second measurement of a different combination of scalar
and tensor polarizabilities.  \drdf{This is accomplished in the
present work by the application of a dc electric field that is
parallel to the standing wave (Fig.~\ref{fig:setup}), whereas the
previous measurement was performed with a dc field that was
perpendicular to the standing wave. In addition, the current
experiment includes a strong magnetic field not present in the
previous case. The magnetic field makes possible the measurement of
the vector polarizability, as discussed in
Section~\ref{sec:Outlook}.}

\subsection{Experimental apparatus and field geometry}
\begin{figure}[tb]\center
\includegraphics[width=\columnwidth]{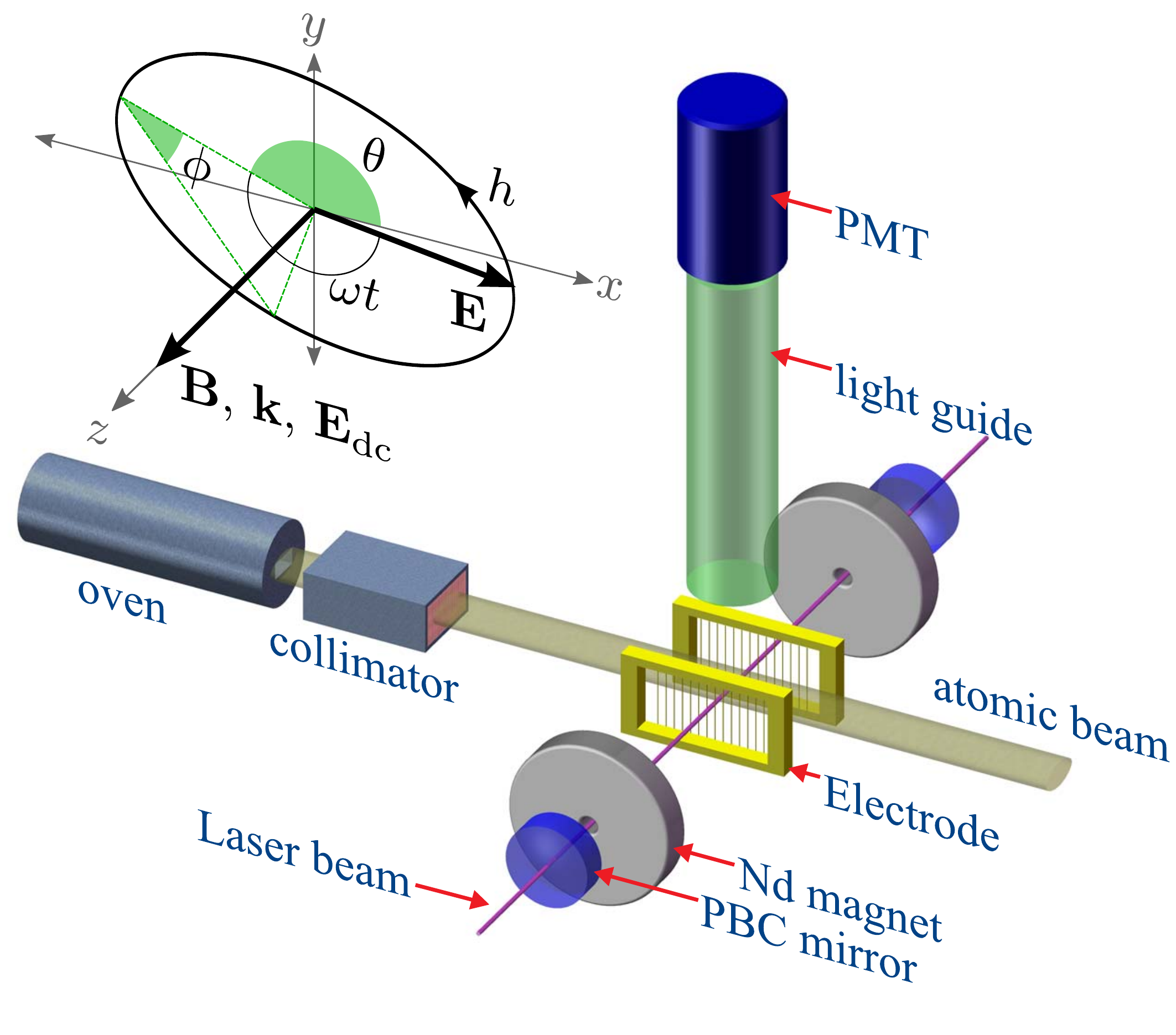}
\caption{\label{fig:setup} \emph{Experimental apparatus.}  A
collimated beam of ytterbium atoms interacts with a standing wave of
light in the presence of dc electric and magnetic fields. The light
is resonant with the 408-nm $\SSzero\rightarrow\TDone$ transition,
and the 556-nm fluorescence is detected by the photomultiplier tube
(PMT). The neodymium (Nd) magnets are axially magnetized in the
$z$-direction. The standing wave is generated in a power buildup
cavity (PBC). With the exception of the PMT, the apparatus is
contained in a vacuum chamber.}
\end{figure}

The details of the experimental apparatus were reported
elsewhere~\cite{ref:Tsigutkin2010}, and only a brief description is
provided here.  A schematic of the setup is shown in
Fig.~\ref{fig:setup}. A beam of Yb atoms is produced  by a
stainless-steel oven loaded with Yb metal, operating at 500
${}^\circ$C. The oven is outfitted with a multislit nozzle, and
there is an external vane collimator reducing the spread of the
atomic beam in the $z$-direction.  Downstream from the collimator,
atoms enter a region with three external fields: a uniform, static
magnetic field \Bfield; a uniform, static electric field \Efield;
and a non-uniform, dynamic electric field~$\Elight(\rvec,t)$.

 % ==== FIGURE: Experimental Setup ====

The magnetic field \Bfield is generated by a pair of axially
magnetized neodymium (Nd) magnets.  These magnets produce a field
with sufficient strength (more than 50~G) to completely isolate the
Zeeman sublevels of the upper state $\ket{\TDone}$. The electric
field \Efield is generated by two wire-frame electrodes separated by
2~cm. The ac electric field $\Elight(\rvec,t)$ is due to
standing-wave light at the transition wavelength of 408.346 nm in
vacuum, which is produced by doubling the frequency of the output of
a Ti:sapphire laser (Coherent 899). About 7 mW of \mbox{408-nm}
light is coupled into a power buildup cavity (PBC) with finesse of
approximately~15,000. The PBC is an asymmetric cavity with a flat
input mirror and a curved back mirror with a 50-cm radius of
curvature. The separation between the mirrors is 22~cm.

Fluorescent light with a wavelength of 556 nm is collected with a
light guide and detected with a photomultiplier tube (PMT).  With
the exception of the PMT, the entire apparatus is contained within a
vacuum chamber with a residual gas pressure of $3\times10^{-6}$
Torr.

As can be seen in Fig.~\ref{fig:setup}, the fields \Bfield and
\Efield point in the \zhat direction.  Likewise, the standing wave
is oriented along the $z$-axis.  The light field $\Elight(\rvec,t)$
lies in the $xy$-plane. For this geometry, the transition to the
upper state $\ket{\TDone;M_a=0}$ is suppressed.

% ==== FIGURE: Energy Level Diagram ====
\begin{figure}[tb]\center
\includegraphics[width=\columnwidth]{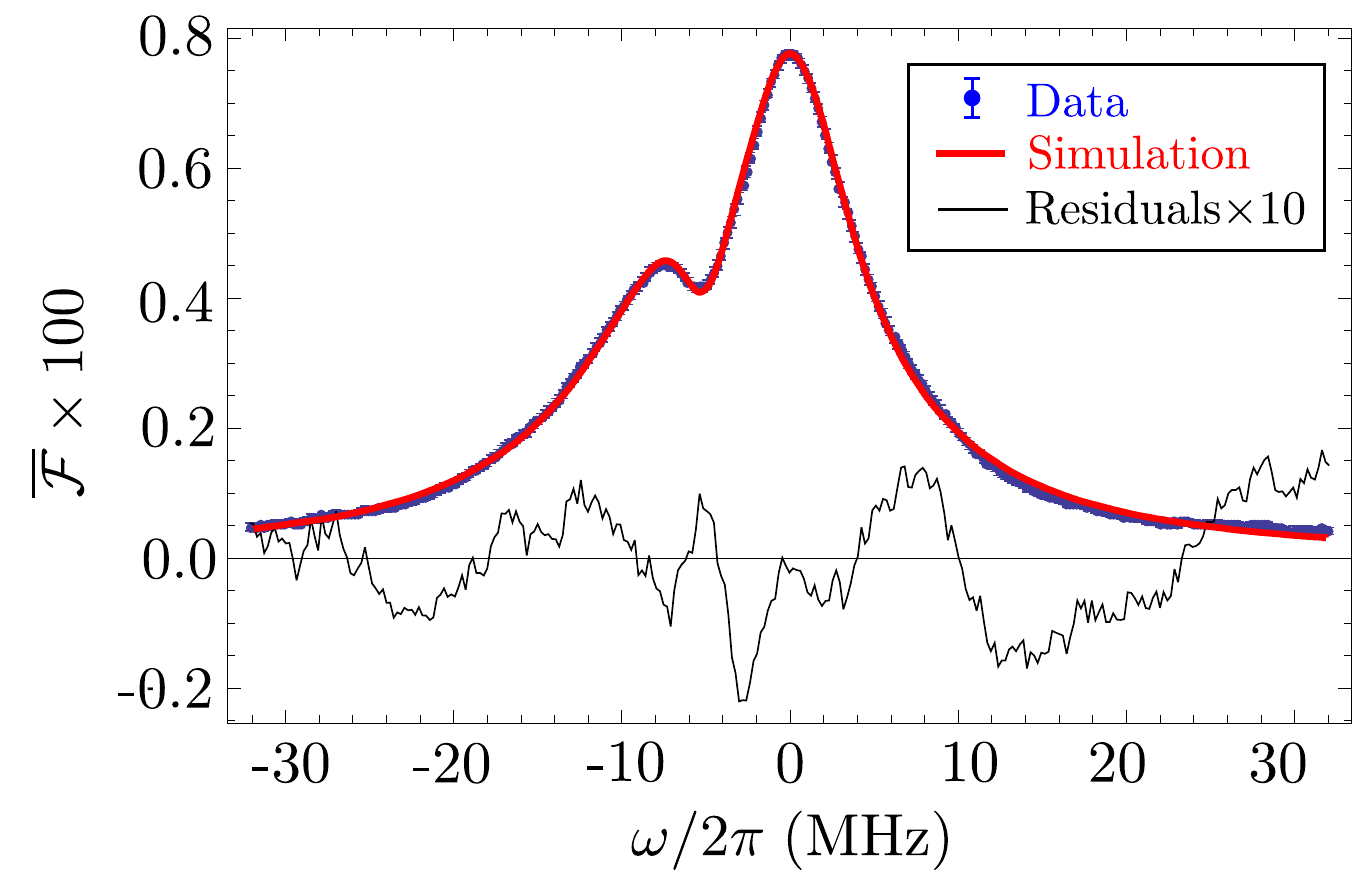}
\caption{\label{fig:FittedLineshape} \emph{Results of fitting
program.} Shown is a comparison of the observed and the simulated
lineshapes of the $\SSzero\rightarrow\TDone$ transition.  The data
correspond to Run 3 in Table \ref{tab:Results}. Also shown are the
residuals, magnified by a factor of 10. The residuals are the
difference of the data and the simulation. }
\end{figure}

The polarization of the light field is of the form \mbox{$\ehat =
\epsilon_x\,\xhat+\epsilon_y\,\yhat$}, where
\mbox{$|\epsilon_x|^2+|\epsilon_y|^2=1$}. To further characterize
the polarization, we introduce three parameters: polarization angle
$\theta$, degree of ellipticity $\phi$, and handedness $h$, which
are given by~\cite{ref:Auzinsh2010}
\begin{align}
\tan2\theta &=2\,\Re(\epsilon_x\epsilon_y^{\ast})/
(|\epsilon_x|^2-|\epsilon_y|^2),\label{eq:theta}\\
\sin2\phi&=2\,|\Im(\epsilon_x\epsilon_y^{\ast})|=
|i(\ehat\times\ehat^{\ast})\cdot\zhat|,\label{eq:phi}
\end{align}
and
\begin{equation}
h=-\mrm{sgn}[i(\ehat\times\ehat^{\ast})\cdot\zhat].\label{eq:h}
\end{equation}
Linearly, circularly, and elliptically polarized light are described
by $\phi=0$, $\phi=\pi/4$, and $0<\phi<\pi/4$, respectively. The
sense of rotation is determined by $h$: left and right-handed
polarizations correspond to $h=+1$ and $h=-1$, respectively.
Substituting $\Efield = \efield\,\zhat$ and \mbox{$\ehat =
\epsilon_x\,\xhat+\epsilon_y\,\yhat$} into Eqs.~(\ref{eq:dagCag})
and (\ref{eq:aac}) yields
\begin{equation}
\dind  = \beta_{ag}\efield\sqrt{(1/2)[1+hM_a\sin(2\phi)]},
\end{equation}
and
\begin{equation}\label{eq:aag_with_phi}
\aac_a = \aac_0(\TDone)-hM_a\sin(2\phi)
\aac_1(\TDone)-\frac{1}{2}\aac_2(\TDone),
\end{equation}
for $M_a=\pm1$. Here we have used Eqs.~(\ref{eq:phi}) and
(\ref{eq:h}) to eliminate $\epsilon_x$ and $\epsilon_y$ in favor of
the degree of ellipticity $\phi$ and the handedness $h$. \drdf{For
this geometry, both $\dind$ and $\aac_a$ are independent of the
polarization angle $\theta$.}

According to the geometry in Fig.~\ref{fig:setup}, the component of
the atom's velocity that is perpendicular to the standing wave is
\mbox{$v_\perp = (v_x^2+v_y^2)^{1/2}$}. The output of the oven is
about 6~mm in the $y$-direction, and is located more than 20~cm away
from the standing wave. In order for an atom to pass through the
standing wave, its velocity components must satisfy $|v_y/v_x|\leq
0.015$. Therefore, the approximation \mbox{$v_\perp= v_x$} is valid
and the use of the thermal distribution given in
Eq.~(\ref{eq:woven}) is justified. However, for a high-precision
measurement, the effect of this approximation needs to be
investigated.

\subsection{Data analysis}

\begin{ruledtabular}
\begin{table}[tb]\center
\caption{\label{tab:Results} \emph{Summary of results.} The
uncertainty in the fitted parameters is the asymptotic standard
error returned by the fitting algorithm.}
\begin{tabular}{c|lllll|l}
Run & $c_0\!\cdot\!10^2$ & $c_1\!\cdot\!10^2$ &
\multicolumn{1}{c}{$S_0$} & \multicolumn{1}{c}{$S_1$} &
\multicolumn{1}{c|}{$S_2$} & \multicolumn{1}{c}{$S_1S_2/S_0$} \\
\hline
1 & 5.1(1) & 5.5(4) & 2.83(17) & -0.208(3) & 34(1.1) &-2.53(16) \\
2 & 5.0(1) & 3.2(3) & 2.76(16) & -0.215(3) & 30(1.1) &-2.33(17) \\
3 & 5.4(1) & 7.0(3) & 2.68(12) & -0.187(2) & 34(1.0) &-2.35(13) \\
4 & 5.0(1) & 4.7(3) & 2.56(17) & -0.195(3) & 29(1.1) &-2.22(17)
\\ \hline Avg.: & & & 2.70(7)  & -0.199(1) & 31.8(5) &-2.36(8)
\end{tabular}
\end{table}
\end{ruledtabular}

The data were acquired over four separate experiments (runs). During
each run, approximately 2000 lineshapes were recorded at a rate of
about 100~ms per lineshape and an average lineshape was computed.
This procedure resulted in 4 lineshapes, each with an effective
integration time of about 200 s. A typical lineshape is shown in
Fig.~\ref{fig:FittedLineshape}. After fitting the theoretical model
to the data, the statistical uncertainties of each run were scaled
to give a reduced $\chi^2$ of unity. The scale factor varied between
4.0 and 4.7 depending on the run. The resulting error bars are shown
in the figure. The relatively large scale factor indicates that the
accuracy of the fit is dominated by either systematic distortion of
the profile during the scan, or profile features that are neglected
in the theoretical model, but not by the statistical uncertainty of
the signal.

In the present experiment, the oven temperature was
\mbox{$T=500(50)\;{}^{\circ}\mrm{C}$} and the magnitude of the dc
electric field was \mbox{$\efield = 4.24(6)$ kV/cm}.  The atoms
intersect the standing wave in the middle of the PBC where the
radius of the light beam is \mbox{$\rbeam=196(5)\;\mathrm{\mu m}$}.
We used linearly polarized light with \mbox{$\phi=0(1)^{\circ}$}.
Then Eqs. (\ref{eq:aacu}) through (\ref{eq:wu}) become
\begin{align}
\aac_{ag}^{\mrm{II}} &= [0.0668(86)\;\mathrm{Hz\cdot(V/cm)^{-2}}]\times S_1S_2/S_0,
\label{eq:aII}\\
P &= [5.02(98)\;\mathrm{W}]\times S_{0},
\end{align}
and
\begin{equation}\label{eq:omegau}
\omegaD = [2\pi\times0.209(17)\;\mrm{MHz}]\times S_2,
\end{equation}
where $\aac_{ag}^{\mrm{II}} =
\alpha_0(\TDone)-(1/2)\alpha_2(\TDone)-\alpha_0(\SSzero)$.  Here the
superscript ``II" is used to distinguish the results of the present
work from the previous measurement.  The dimensionless decay rates
are \mbox{$G_{ab}=0.35(4)$}, \mbox{$G_{b}=0.45(4)$}, and
\mbox{$G_c=0$}, and the parameter $S_3$ is given by
\mbox{$S_3=0.53(5)$}.

As part of our analysis, we normalized the observed and simulated
lineshapes by their maximum values.  In addition, we introduced two
calibration parameters to the simulated curve:
\begin{equation}
\overline{\mathcal{F}}(x;\mbf{S},\mbf{G},\mbf{c})=
(1+c_0)\overline{\mathcal{F}}((1-c_1)x;\mbf{S},\mbf{G})-c_0,
\end{equation}
where $\mbf{c}=(c_0,c_1)$.  Here $c_0$ accounts for the background
of the observed signal. The parameter $c_1$ is a scaling factor that
accounts for any variation in the calibration of the frequency axis
of the data relative to the simulation. Such deviations could arise
due to misalignment of the atomic beam relative to the axis of the
PBC, deviations in the perpendicular velocity distribution, or the
uncertainty of the timescale $1/\Gamma_a$.  In practice, both $c_0$
and $c_1$ represent very small corrections, with typical values on
the order of $0.05$ (see Table~\ref{tab:Results}).

The results of the fitting for each run are given in
Table~\ref{tab:Results}.  Combing these results with
Eqs.~(\ref{eq:aII}) through (\ref{eq:omegau}) yields
\begin{equation}
\aac_{ag}^{\mrm{II}} = -0.158(21)\;\mrm{Hz\cdot(V/cm)^{-2}},
\end{equation}
$P =13.6(2.6)\;\mrm{W}$, and $\omegaD =
2\pi\times6.58(53)\;\mrm{MHz}$. \drdf{The measured values of the
circulating power $P$ and the Doppler broadening $\omegaD$ are
consistent with (and more precise than) direct measurements of these
quantities.} A comparison of the data to the fit is shown in
Fig.~\ref{fig:FittedLineshape}.  {The quality of both the data and
the fit are sufficiently high that the uncertainty of the measured
value of $\aac_{ag}^{\mrm{II}}$ is primarily due to the
uncertainties in the vector transition polarizability $\beta_{ag}$
and the linewidth $\Gamma_a$. A summary of the factors that
contribute to the uncertainty of the measurement are shown in
Table~\ref{tab:ErrorBudget}.}

\begin{ruledtabular}
\begin{table}[tb!]\center
\caption{\label{tab:ErrorBudget} \emph{Error budget.} Shown are the
factors contributing to the uncertainty of the measured value of
$\aac_{ag}^{\mrm{II}}$.}
\begin{tabular}{lr}
Factor & Uncertainty (\%) \\ \hline
Vector transition polarizability ($\beta_{ag}$) & 9\\
Lifetime of \TDone & 8\\
Light polarization & 4\\
DC electric field & 3\\
Simulation and data fit & 3 \\
\hline {\bf Total} (in quadrature) & {\bf 13}
\end{tabular}
\end{table}
\end{ruledtabular}

The quantity $\aac_{ag}^{\mrm{II}} =
\alpha_0(\TDone)-(1/2)\alpha_2(\TDone)-\alpha_0(\SSzero)$ is a
combination of the ac scalar and tensor polarizabilities of the
states \TDone and \SSzero.  The tensor polarizability of \TDone is
determined unambiguously by comparing the present measurement of
$\aac_{ag}^{\mrm{II}}$ with the previous measurement of
$\aac_{ag}^{\mrm{I}}$, which was discussed after Eq.~(\ref{eq:aI}).
We find
\begin{equation}
\alpha_2(\TDone) = -0.103(26)\;\mrm{Hz\cdot(V/cm)^{-2}},
\end{equation}
and
\begin{equation}\label{eq:a0difference}
\alpha_0(\TDone)-\alpha_0(\SSzero) =
-0.209(18)\;\mrm{Hz\cdot(V/cm)^{-2}}.
\end{equation}
Finally, the scalar polarizability of \TDone is isolated by
substituting the calculated value $\aac_0(\SSzero)$, given by
Eq.~(\ref{eq:alpha0_estimate}), into Eq.~(\ref{eq:a0difference}):
\begin{equation}
\alpha_0(\TDone) = 0.009(21)\;\mrm{Hz\cdot(V/cm)^{-2}}.
\end{equation}
With the present accuracy, the scalar polarizability of \TDone is
consistent with zero.  In the presence of linearly polarized light,
the polarizability of \TDone is dominated by the tensor
polarizability. However, for light with arbitrary polarization,
$\alpha_1(\TDone)$ also plays a role.  The vector polarizability is
the subject of ongoing experiments.

% =====================================
% OUTLOOK
% =====================================
\section{Summary and Outlook}\label{sec:Outlook}

This work is part of a continuing investigation of polarizabilities
in Yb. The ac scalar and tensor polarizabilities of the excited
\TDone state in Yb were measured independently for the first time.
Ongoing experiments are focused on measuring the ac vector
polarizability, for which there is currently no experimental or
theoretical data.

To measure the vector polarizability, the $\SSzero\rightarrow\TDone$
transition must be excited using circularly polarized light, as can
be seen in Eq.~(\ref{eq:aag_with_phi}).  Such a measurement requires
control over the ellipticity of the light. In the present
experimental setup, only two Zeeman sublevels ($M_a=\pm1$) are
excited.  The degree of ellipticity can be measured by comparing the
relative strengths of the transitions to different sublevels. For
purely circularly polarized light, only one sublevel is excited.
This condition is ideal for measurement of the vector
polarizability.

In this paper, we presented the next generation of the Lineshape
Simulation Method (LSM) for measuring combinations of
polarizabilities of the ground and excited states in atoms.  The LSM
was originally developed specifically for Yb, but we have
generalized the method for an arbitrary atomic system with the level
structure shown in Fig.~\ref{fig:FittedLineshape}.  For example, the
LSM could be used to measure the polarizabilities of the $6S$ and
$7S$ states in cesium by observing the lineshape of the 539-nm
$6S\rightarrow 7S$ transition driven by a standing wave of
light~\cite{ref:Wieman1987}.

\section{Acknowlegements}
The authors acknowledge helpful discussions with and important
contributions of S.~Corinaldi, A.~Derevianko, V.~A.~Dzuba,
N.~A.~Leefer, S.~M.~Rochester, and J.~E.~Stalnaker. This work has
been supported by NSF.

\appendix

\section{Frequency dependence of dynamic
polarizabilities}\label{app:A} The ac polarizability $\aac_\ell$ of
the state $\ket{\ell}=\ket{\gamma J M}$ is given by
Eq.~(\ref{eq:aac}). The scalar, vector, and tensor polarizabilities
depend on the light frequency $\omega$ in the following
way~\cite{ref:Bonin1994,ref:Bonin1997}:
\begin{align}
\aac_{0}(\gamma J)&=\frac{1}{3\hbar}\sum_{\ell'\neq
\ell}\frac{|d_{\ell\ell'}|^2}{2J+1}\,F^{+}_{\ell\ell'}(\omega)\;
\Phi_{0}(J,J'),\label{eq:alpha0}\\
\aac_1(\gamma J) &= \frac{1}{2\hbar}\sum_{\ell'\neq
\ell}\frac{|d_{\ell\ell'}|^2}{2J+1}\,F^{-}_{\ell\ell'}(\omega)
\;\Phi_1(J,J'),\\
\aac_{2}(\gamma J)&=\frac{1}{3\hbar}\sum_{\ell'\neq
\ell}\frac{|d_{\ell\ell'}|^2}{2J+1}\,
F^{+}_{\ell\ell'}(\omega)\;\Phi_{2}(J,J').\label{eq:alpha2}
\end{align}
Here the summation is over all states
$\ket{\ell'}=\ket{\gamma'J'M'}$ such that $\ket{\ell}$ and
$\ket{\ell'}$ have opposite parity. The functions $\Phi_0$, $\Phi_1$
and $\Phi_2$ are given by
\begin{align}
\Phi_0(J,J') &= \delta_{JJ'}+\delta_{J,J'+1}+\delta_{J,J'-1},\\
\Phi_1(J,J') &=
-\frac{\delta_{JJ'}}{(J+1)}+\frac{J\,\delta_{J,J'+1}}{J+1}
-\delta_{J,J'-1},
\end{align}
and
\begin{equation}
\Phi_2(J,\!J')\!=\!\frac{(2J\!-\!1)\delta_{JJ'}}{J\!+\!1}\!-
\!\frac{J(2J\!-\!1)\delta_{J,J'\!+\!1}}{(J\!+\!1)(2J\!+\!3)}\!-
\!\delta_{J,J'\!-\!1},
\end{equation}
where $\delta_{JJ'}$ is the Kronecker delta. We emphasize that
$\Phi_1(J,J')$ differs from the expression found in
Refs.~\cite{ref:Bonin1994,ref:Bonin1997} by a factor of $J$.  The
reason for this discrepancy is that we follow the convention for
which the vector polarizability of a stretched state ($M=J$) is
$\alpha_1(\gamma J)$ instead of $J\times\alpha_1(\gamma J)$.

The frequency-dependent parts of Eqs.~(\ref{eq:alpha0}) through
(\ref{eq:alpha2}) are given by
\begin{equation}\label{eq:Fpm}
F^{\pm}_{\ell\ell'}(\omega) = \frac{1}{\omega_{\ell'
\ell}-\omega-i\Gamma_{\ell'}/2}\pm \frac{1}{\omega_{\ell'
\ell}+\omega+i\Gamma_{\ell'}/2}.
\end{equation}
Here \mbox{ $d_{\ell\ell'}\equiv\reducedME{\gamma
J}{d}{1}{\gamma'J'}$} is the reduced matrix element of the electric
dipole operator and $\kq{d}{1}{}$ is the spherical tensor associated
with the electric dipole moment. In the limit of large detuning from
resonance $(|\omega_{\ell' \ell}\pm\omega|\gg\Gamma_{\ell'})$, the
functions $F^{\pm}_{\ell\ell'}$ reduce to
\begin{equation}
F_{\ell\ell'}^+ = \frac{2\omega_{\ell' \ell}}{(\omega_{\ell'
\ell}^2-\omega^2)} \quad\mrm{and}\quad F_{\ell\ell'}^- =
\frac{2\omega}{(\omega_{\ell' \ell}^2-\omega^2)}.
\end{equation}

\section{System of equations used in numerical model}\label{app:B}
For computational purposes, it is convenient to express
Eqs.~(\ref{eq:sigma}) and (\ref{eq:Flo}) as
\begin{equation}\label{eq:OBE2}
\frac{d\boldsymbol{\rho}}{d\tau} =
\mathbf{f}(\boldsymbol{\rho},\tau), \quad
 \boldsymbol{\rho}(-\tau_0)= 0,
\end{equation}
where $\tau_0=3/(S_3u_{\perp})$.  Here \mbox{$\boldsymbol{\rho} =
(\rho_0,\hdots,\rho_5)$}, \mbox{$\rho_0=\rho_{aa}$},
\mbox{$\rho_1=\rho_{bb}$}, \mbox{$\rho_2=\rho_{cc}$},
\mbox{$\rho_3=\Re[\rho_{ga}]$}, \mbox{$\rho_4= \Im[\rho_{ga}]$},
\mbox{$\rho_5=\mathcal{F}$}, and \mbox{$\mathbf{f} =
(f_0,\hdots,f_5)$}. The components of $\mathbf{f}$ are given by
\begin{align}
f_0&=-\rho_0 +\Gammap\,\rho_4,\\
f_1&=-G_b\,\rho_1 + G_{ab}\,\rho_0,\\
f_2&=-G_c\,\rho_2+G_{ac}\,\rho_0,\\
f_3&=-\frac{1}{2}\rho_3-\Delta\rho_4,\\
f_4&=-\frac{1}{2}\rho_4+\Delta\rho_3 -\frac{\Gammap}{2}
(2\rho_0\!+\!\rho_1\!+\!\rho_2\!-\!1),\\
f_5&=+G_b\,\rho_1,
\end{align}
where $\Gammap$ and $\Delta$ are given by Eqs.~(\ref{eq:Gamma_pump})
and (\ref{eq:Delta}), respectively. To derive Eq. (\ref{eq:OBE2}),
we eliminated the population $\rho_{gg}$ of the ground state from
the OBE using the conservation of probability:
\mbox{$\sum_{n}\rho_{nn}=1$}. The fluorescence defined in
Eq.~(\ref{eq:Flo}) is given by
\begin{equation}
\mathcal{F}(\boldsymbol{\xi},\mbf{u}) =
\rho_5(\tau_0;\boldsymbol{\xi},\mbf{u}).
\end{equation}
Thus the fluorescence can be obtained by numerically solving the
system of equations (\ref{eq:OBE2}), as described in the text.

\end{document}